\documentclass[a4paper, 11pt, oneside]{article}
\usepackage[T1]{fontenc}
\usepackage{geometry}
\usepackage{amsthm}
\usepackage{amssymb}
\usepackage{mathdots}
\usepackage[longnamesfirst]{natbib} 
\usepackage[dvipdfmx,hiresbb]{graphicx} 
\usepackage{array,booktabs} 
\usepackage{setspace} 
\usepackage{multirow} 
\usepackage{threeparttable} 
\usepackage{lscape} 
\usepackage{hyperref} 
\usepackage{endfloat} 
\newcommand{\source}[1]{\caption*{\small Source: {#1}} } 
\newenvironment{ltable} 
{\begin{landscape}\vspace*{1.5cm}\begin{table}}
		{\end{table}\end{landscape}}
\DeclareDelayedFloatFlavor{ltable}{table}
\makeatletter

\usepackage{subcaption} 
\usepackage{comment} 
\usepackage{color} 
\usepackage{mathpazo} 
\begin{document}

\title{Quantifying Health Shocks Over the Life Cycle}

\author{Taiyo Fukai\thanks{We thank Professors Hideo Yasunaga, Hiroki
        Matsui, and Yusuke Sasabuchi for letting us use their data and
        for answering our numerous questions while we were preparing for this paper.  We also thank
        Dr.Ilja Musulin for his editing work and extend our gratitude to the Japan Society for the Promotion of Science (JSPS)
        for its Grant-in-Aid for Scientific Research 15H05692, as well as to the Research Institute of Economy, Trade and Industry (RIETI) for their
        contribution to the Japanese Study of Aging and Retirement (JSTAR).}\footnote{fukai-taiyo@g.ecc.u-tokyo.ac.jp}, Hidehiko Ichimura\footnotemark[1]\footnote{ichimura@e.u-tokyo.ac.jp},
        Kyogo Kanazawa\footnotemark[1]\footnote{kanazawa-kyogo285@g.ecc.u-tokyo.ac.jp}\\ Graduate School of Economics,
        University of Tokyo}

\date{August 2, 2017}

\maketitle

%
\begin{abstract}
 We first show (1) the importance of investigating health expenditure process using the order two Markov
 chain model, rather than the standard order one model,
 which is widely used in the literature.  Markov chain of order two is the
 minimal framework that is capable of distinguishing those who
 experience a certain health expenditure level for the first time from those
 who have been experiencing that or other levels for some time.
 In addition, using the model we show (2) that the probability of encountering a
 health shock first decreases until around age 10, and then increases with age, particularly, after age 40, (3) that health shock distributions
 among different age groups do not differ until their percentiles reach the median
 range, but that above the median the health shock
 distributions of older age groups gradually start to first-order dominate those
 of younger groups, and (4) that the persistency of health shocks also shows a U-shape in relation to age. (\textit{JEL}: I10, I12, I18)
\end{abstract}

\doublespacing

\section{Introduction}
How one's health may evolve conditional on the current health status as
one ages is a major concern for anyone.  It is also an important
element in individuals' decision making regarding savings.  From the perspective of private
firms, it is a significant human resource management issue, and for
providers of private health and life insurances, a crucial component in pricing their insurance policies competitively.  
From the viewpoint of public policy, understanding this process is vital for designing socially desirable health care and health insurance
policies.  The process of health state transition is also an important constituent in
various economic models that are used to evaluate alternative social
policies.

This paper attempts to shed light on the health expenditure process by taking the incurred annual
medical cost as representing the health status, and by examining the
information on medical costs found in receipts stored in a database which
contains observations on over three million individuals for the period of up to 11 years.

We use the age-dependent Markov chain of order two to capture the health
state transition instead of the usual Markov chain of order one.  This
allows us to identify states with different levels of health shocks
by taking into consideration states with a zero expenditure level in the previous year and a
certain expenditure level in the current year.  Markov chain of order
one cannot isolate such states.

The four main findings of the present paper are as follows. First, conditioning only 
on the state from a previous year cannot predict a patient's future health expenditure path
to the extent that conditioning on the state from two previous years can.
Second, the probability of encountering a health shock first decreases until around age 10, and then increases with age, particularly, after age 40.
Third, health shock distributions among different age groups do not differ until their percentiles reach the median range, but, above the median, the health shock distributions of older age groups gradually start to first-order dominate those of younger groups.
And fourth, the persistency of health shocks also shows a U-shape in relation to
age.

We describe the methodological framework we use in section 2.  After describing the
institutional background in Japan and the database we utilized in section 3, in section 4 we present our estimation method. 
The main findings of the paper are discussed in section 5, while section 6 contains a conclusion.

\section{Related Literature}
Since long ago, the persistency of medical expenditure in individuals, the phenomenon that those who have a high health expenditure in one year
tend to continue paying high medical fees in the following years too, has been observed and studied.
For example, studies by \cite{mccallwai}, \cite{andersonknickman}, \cite{beebe} and \cite{freebornetal} reveal such persistency among Medicare beneficiaries, while
\cite{newhouse}, \cite{vliet} and \cite{coulsonstuart} point to a positive correlation between high expenditure in adjacent years 
and a positive but relatively low correlation between high expenditure in years farther apart, using data from other medical insurance schemes.
More recently, a number of studies have divided the population based on the quantiles of expenditure incurred in a certain year and
traced the expenditure transition following that year. 
\citep{garber, monheit, paulyzeng, riley, cohenyu, hirthetal1, hirthetal2, karlssonetal}.
However, this type of analysis becomes problematic when we compare the results for different age groups, 
and our approach of defining patient categories based on the subjects' absolute expenditure overcomes that problem - a point we will elaborate on in more detail later.
\cite{eichneretal} is an exceptional study in that it uses the absolute value of health expenditure to divide the population in addition to the quantiles.
Furthermore, \cite{rettenmaierwang} conduct an estimation of Medicare reimbursement by using the Tobit model and show that the lag variable has a statistically significant positive effect.
Also, \cite{kohnliu} use a British dataset on medical care use, and not health expenditure, and observe a similar persistency in the utilization of that care.
As for studies using Japanese data, \cite{kansuzuki} find a stronger health expenditure persistence than that observed in research using U.S. data.
In addition, \cite{Suzukietal} as well as \cite{Ibukaetal} conduct the same kind of analysis as mentioned above - using quantiles of expenditure in a certain year and observing the transition into the following year, while \cite{masuhara} follows the analyses in \cite{eichneretal}.  

Health expenditure dynamics are also investigated by using continuous health expenditures.
For example, \cite{feenbergskinner} study the time-series properties of health expenditures by utilizing the health expenditure panel data provided by the U.S. Internal Revenue Service (IRS).
They use an ARMA model and find that the ARMA(1, 1) model better fits the covariance structure of medical expenses, suggesting the persistency of medical expenditures.
Following \cite{feenbergskinner}, and using the Health and Retirement Survey (HRS) and the Assets and Health Dynamics of the Oldest Old (AHEAD), \cite{frenchjones} also find a highly persistent AR(1) health expenditure process.
However, while their approach can capture well the fact that health expenditures persist, their model specifications, such as the constant persistency parameter, possibly miss heterogeneity in age and the initial health status, as we shall later discuss.

The literature volume is large and, of course, the methods and results are varied, but what can be said with relative certainty is that there is some persistency in individual health spending 
and that the persistency is not so strong as to enable us to predict future expenditure with high accuracy based solely on the expenditure in a certain year.
For example, \cite{hirthetal1} use U.S. private insurance data for subjects under age 65 from 2003 to 2006 
to show that 43\% of enrollees in the top decile of health spending in a given year remained in the top decile the following year
and that 34\% of them remained in the top decile five years later.
Obviously, this result does indicate persistency, since 43\% and 34\% are far above 10\%, 
but it also suggests that more than a half of the subjects in the top decile left it during the following year.
Our approach, conditioning on the states from two previous years, will help alleviate this problem, by enabling a distinction between individuals in the same expenditure group 
whose health is ``temporarily'' bad and those who suffer from a  ``continuous'' bad condition. 

When it comes to age, some of the previous studies above, including \cite{kansuzuki}, \cite{cohenyu} and \cite{kohnliu}, state that
older people tend to exhibit stronger persistency in health expenditure than the young (\cite{hirthetal2} , however, present the opposite result).
Our rich dataset, however, allowed us to calculate the persistency for each age and thus enabled us to see the differences between age groups in more detail than those studies - we have found that persistency in relation to age exhibits a U-shaped curve.

In addition, there is another strain of literature that attempts to capture individuals' health expenditure through stochastic dynamic individual models.
Some of these studies define expenditure as a variable endogenously determined by individuals. This is the case with 
the health asset model from the classic work of \cite{Grossman}, as well as with some recent studies, such as \cite{HJ} and \cite{Yogo}.
Furthermore, some studies define medical expenses as exogenous expenditure, e.g. \cite{HSZ}, \cite{Palumbo}, \cite{chouetal} and \cite{Capatina}.
However, the solutions of the individual's maximization problem in these studies depend only on the individual's current health status, which is why the authors were only able to describe health expenditure with a Markov chain of order one.
In contrast, our method treats that expenditure as a Markov chain of order two, and thus possesses more predictive power regarding individuals' future expenditure path.

The future elderly model, which was developed by \cite{goldmanetal} and applied to Japanese data in \cite{chenetal} 
is another recent method for predicting individuals' future health expenditure paths based on their current status.
However, this method too can only address health expenditure as a Markov chain of order one, and cannot distinguish between different individuals 
with the same current status.

\section{Methodological Framework}
In the present study we use an age-specific Markov chain of order two as the main analytical
framework. The subjects' health status is defined by the level of incurred medical cost.  
As explained in the previous section, the existing studies have mainly used an
age-dependent Markov chain of order one, where health status is defined by
age-dependent percentiles of health expenditure.  

First, using Markov chain of order two allows us to identify states with
different levels of health shocks by examining the states with zero
expenditure level in the previous year and a certain expenditure level in the
current year.  Markov chain of order one, however, does not let us distinguish between such states.

Second, defining the states by using the absolute health expenditure allows us to
easily compare across ages the changes in the health transitions matrix.
When the states are defined using age-specific percentiles, this
cannot be done easily.

Third, the large sample size of our data set allowed us to estimate the
transition matrix for each age without any parametric assumptions. This is in contrast with some
previous studies, such as \cite{SSK}, \cite{HHL} or \cite{Capatina},
which assumed a functional form, such as polynomials, in age.  Our large sample made it possible for us
to obtain results which might have been difficult to obtain with a parametric
approach.


\section{Background and Data}
In this section, we briefly describe the main features
of the Japanese health care system and the data set used.

\subsection{Background}

Japan has a universal, public health insurance system. In effect, this means that a significant proportion of the medical costs incurred by the users, including prescription drugs, is covered by their health insurance.   
The exact proportion of the cost that the health insurance user is expected to bear varies in accordance with age and income.  
In the current system, most users, aged 7 to 69, pay a 30\% co-payment, whereas those between ages 70 and 74 are entitled to a lower rate of 20\%, unless they possess income comparable to that of the active work force, in which case they too are obliged to contribute 30\% of the incurred health cost.  Those over 75 years of age are required to make a small contribution of 10\%, unless they continue to obtain a significant income that is similar to that of the current workforce. 
For children under 7, the co-payment rate is set at 20 percent.\footnote{It is worth noting that some local governments in Japan offer free medical services to children through subsidy programs aimed at providing support for child-rearing.}

Also, it should be noted that, as a part of the current public health insurance scheme, the users in Japan are spared from paying dramatically high medical expenses. 
The system that makes that possible is called the high-cost medical expense benefit system (or ``\textit{kogaku ryoyohi seido}''in Japanese). Under it, the users are reimbursed a portion of the amount they paid through co-payment in case the cost that they had incurred had been very high.  
That is, the system is designed to keep patients' financial burden relatively light by compensating them if their expenditure exceeds a specified threshold amount.  
The amount depends on the user's age and income level. 
For example, the threshold for users aged 7 to 69 who fall into the medium-income bracket is 80,100 yen per month.

This benefit system is provided by several health schemes, including the National Health Insurance (``\textit{kokumin kenko hoken}'' in Japanese) and the National Federation of Health Insurance Societies (``\textit{kenko hoken kumiai}'' in Japanese), which offers employment-based health insurance.
Under these schemes, insurance users can choose medical institutions freely. 
The National Health Insurance is mostly utilized by persons in agriculture, family business and self-employment, while paid company employees tend to use employment-based schemes provided by the National Federation of Health Insurance Societies (NFHIS).
While insurance fees vary across schemes depending on income and family composition, the content of the medical service covered tends to be basically the same.

\subsection{Data Description}

In this study we employ data on medical insurance claims obtained from the Japan Medical Data Center (JMDC).  
The JMDC claim database contains data on monthly receipts for more than three million people who are covered by employment-based health insurance.
It is a longitudinal database that follows individuals as long as they participate in the same health insurance.  
Company employees as well as their dependents are covered by this type of health insurance.  
We use the JMDC data from fiscal year 2005 to 2015.
As of March 2016 (the end of the FY 2015), the database contained input from more than 90 employment-based health insurance schemes.

The database provides demographic information, such as sex and age of the subjects and their medical costs, including both expenses for treatment and pharmacons.
The receipt data are renewed every month.
Based on that information on monthly medical costs, we calculate the annual expenditure by multiplying the average monthly medical cost with 12. 
This approach is taken because some individuals are in the database only for a part of the year.

It is worth noting that our data set has both advantages and disadvantages.
First, since the JMDC claim database contains only information on receipts, we cannot utilize the socio-economic status of health insurance users, such as their income and educational background.
Also, as mentioned above, not all individuals are present in the database the whole year - some of them drop out because they have withdrawn from the insurance scheme they used and some because they have passed away. 
Unfortunately, we cannot clearly identify the reasons behind the sample attrition.
Nonetheless, when looking at the medical costs just before the subjects' omission, we did not notice any strong evidence that individuals were systematically dropping out from our sample, which is why we believe that the problem caused by sample attrition in our dataset is not so serious.
While we recognize these disadvantages, we wish to point out the tremendous benefit that the precise information on medical expenditure and the large sample size give us.
Our receipt dataset provides information that is much more accurate than the self-reported information on medical expenditure collected by the Health and Retirement Survey (HRS) or the Assets and Health Dynamics of the Oldest Old (AHEAD) , which are often used for examining the process of health transition.  
Also, our large sample size and the longitudinal quality of the dataset are sufficient to enable us to examine the health state transition process by age without making any parametric assumptions.

\begin{figure}[!ht]
	\centering
	\includegraphics[scale=0.11]{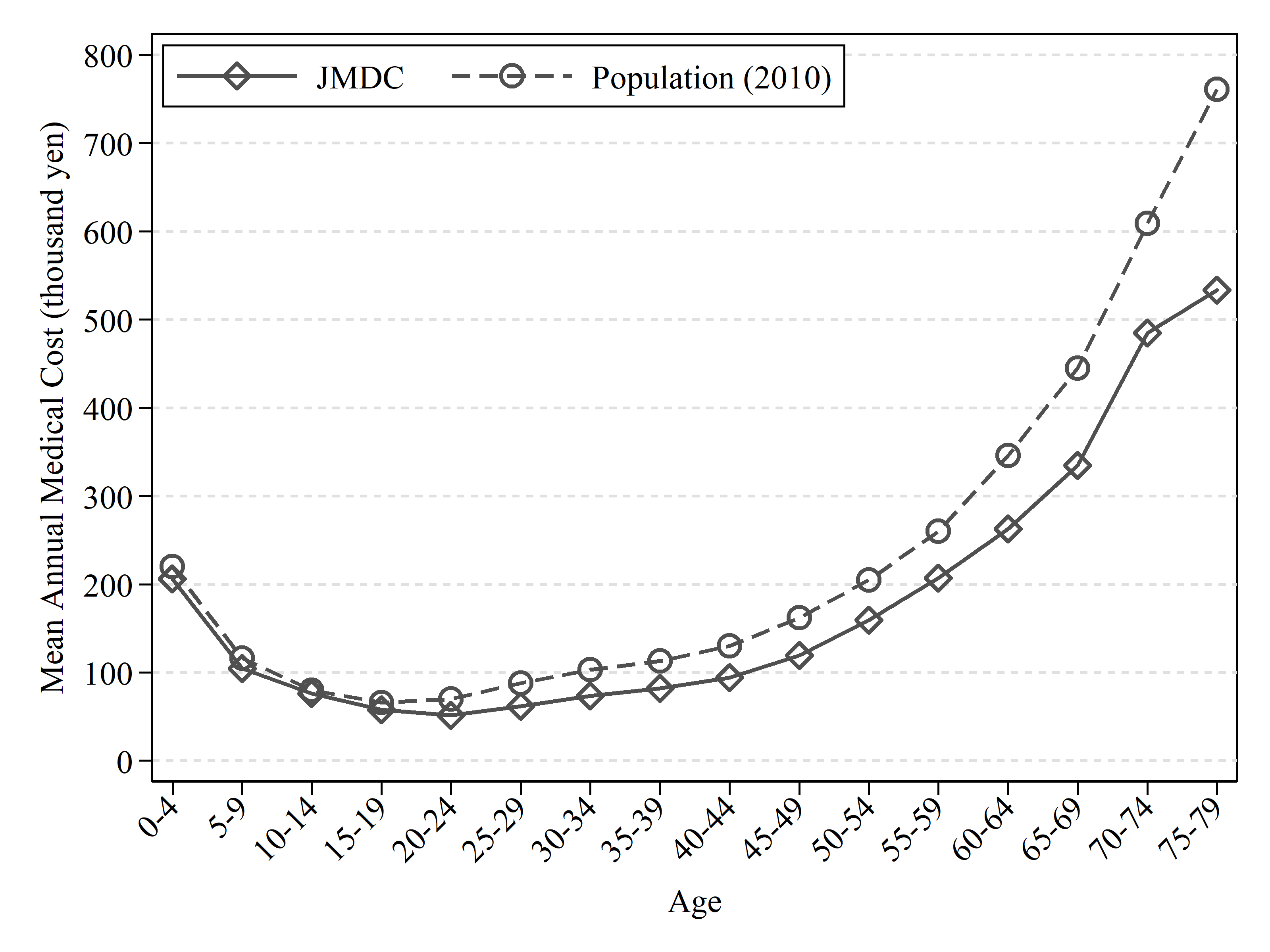}
	\caption{Comparison with the Population Mean} 
	\source{The Japan Medical Data Center (JMDC) claim database and the 2012 annual report "Basic Data on Medical Insurance" by \textit{the Japanese Ministry of Health, Labour and Welfare} (\url{http://www.mhlw.go.jp/file/06-Seisakujouhou-12400000-Hokenkyoku/kiso22.pdf})}
	\label{f01}
\end{figure}

Before discussing the descriptive characteristics of medical costs found in the JMDC claim database, it is important to check how the subjects appearing in that database differ from the overall population.  
First, Table \ref{tab:table1} shows the proportion of individuals who are company employees.  From the table, we can see that almost all the men in their prime working age are insured as company employees. 
However, that is not the case with the women in the database: about two-thirds of them are insured as dependents.  
Second, in comparison with the population mean for medical costs in 2010, found in a report by the Japanese Ministry of Health, Labour and Welfare, the JMDC claim database seems to contain data on relatively healthier people.  
This could be a bias that exists because those who cannot work due to poor health have not been included in the database.  
However, the mean of medical costs for individuals below age 15, who cannot be employees since they are under age minors, seems to provide assurance that the JMDC claim database is representative, since the data for that age group come across as similar to the population mean.  
In addition, the medical fee profiles share similar trends, as indicated in Figure \ref{f01}.

\begin{table}[htbp]
	\centering
	\begin{threeparttable}
	\caption{Fraction of Self-insured Individuals in the JMDC Claim Database and their Mean Annual Medical Costs (thousand yen) in Comparison with the Population Mean}
	\begin{tabular}{rrccrcc}
		\toprule
		\midrule
		&       & \multicolumn{2}{c}{Fraction of Self-insured} &       & \multicolumn{2}{c}{Medical Costs (mean)} \\
		\cmidrule{3-4} \cmidrule{6-7}
		\multicolumn{1}{c}{Age} & \multicolumn{1}{c}{} & Male  & Female & \multicolumn{1}{c}{} & \multicolumn{1}{c}{Population} & \multicolumn{1}{c}{JMDC} \\
		\midrule
		\multicolumn{1}{c}{0-4} & \multicolumn{1}{c}{} & 0.00\% & 0.00\% & \multicolumn{1}{c}{} & 220   & 206.0 \\
		\multicolumn{1}{c}{5-9} & \multicolumn{1}{c}{} & 0.00\% & 0.00\% & \multicolumn{1}{c}{} & 116   & 104.6 \\
		\multicolumn{1}{c}{10-14} & \multicolumn{1}{c}{} & 0.00\% & 0.00\% & \multicolumn{1}{c}{} & 80    & 76.2 \\
		\multicolumn{1}{c}{15-19} & \multicolumn{1}{c}{} & 10.39\% & 3.42\% & \multicolumn{1}{c}{} & 66    & 57.6 \\
		\multicolumn{1}{c}{20-24} & \multicolumn{1}{c}{} & 58.20\% & 37.96\% & \multicolumn{1}{c}{} & 70    & 51.4 \\
		\multicolumn{1}{c}{25-29} & \multicolumn{1}{c}{} & 94.03\% & 59.93\% & \multicolumn{1}{c}{} & 88    & 61.8 \\
		\multicolumn{1}{c}{30-34} & \multicolumn{1}{c}{} & 98.61\% & 43.03\% & \multicolumn{1}{c}{} & 103   & 73.3 \\
		\multicolumn{1}{c}{35-39} & \multicolumn{1}{c}{} & 99.47\% & 34.04\% & \multicolumn{1}{c}{} & 113   & 81.9 \\
		\multicolumn{1}{c}{40-44} & \multicolumn{1}{c}{} & 99.71\% & 29.93\% & \multicolumn{1}{c}{} & 130   & 94.0 \\
		\multicolumn{1}{c}{45-49} & \multicolumn{1}{c}{} & 99.79\% & 27.69\% & \multicolumn{1}{c}{} & 162   & 119.2 \\
		\multicolumn{1}{c}{50-54} & \multicolumn{1}{c}{} & 99.75\% & 26.08\% & \multicolumn{1}{c}{} & 205   & 159.6 \\
		\multicolumn{1}{c}{55-59} & \multicolumn{1}{c}{} & 99.45\% & 24.77\% & \multicolumn{1}{c}{} & 260   & 206.8 \\
		\multicolumn{1}{c}{60-64} & \multicolumn{1}{c}{} & 98.19\% & 22.71\% & \multicolumn{1}{c}{} & 346   & 262.8 \\
		\multicolumn{1}{c}{65-69} & \multicolumn{1}{c}{} & 95.01\% & 13.64\% & \multicolumn{1}{c}{} & 445   & 334.8 \\
		\multicolumn{1}{c}{70-74} & \multicolumn{1}{c}{} & 81.70\% & 5.44\% & \multicolumn{1}{c}{} & 609   & 485.0 \\
		\multicolumn{1}{c}{75-79} & \multicolumn{1}{c}{} & 68.58\% & 4.37\% & \multicolumn{1}{c}{} & 761   & 533.5 \\
		\midrule
		\bottomrule
	\end{tabular}
	\begin{tablenotes}
		\small
		\item Data Sources: The Japan Medical Data Center (JMDC) claim database and the 2012 annual report "Basic Data on Medical Insurance" by \textit{the Japanese Ministry of Health, Labour and Welfare} (\url{http://www.mhlw.go.jp/file/06-Seisakujouhou-12400000-Hokenkyoku/kiso22.pdf})
	\end{tablenotes}
	\label{tab:table1}%
	\end{threeparttable}
\end{table}

Summary statistics are presented in Table \ref{tab:table2}.  
Means of medical costs show a U-shaped age profile, with the bottom around the age of 20.  
The standard deviation of medical costs, too, exhibits a U-shaped age profile.
That is, children and old people tend to incur volatile or extremely high medical costs.  
Furthermore, when we look at the distributional characteristics of the medical costs, we first notice that the mean and the median values are largely different (Table \ref{tab:table2}).  
This is because the upper tail (95 or 99 percentiles) is far longer than the lower tails (5
or 1 percentiles). 
In addition, the upper tail becomes longer as one grows older.  
We also find that there are about 10\% of those who did not receive any medical services for the period of one year.
\begin{ltable}[htbp]
		\centering
		\begin{threeparttable}
		\caption{Summary Statistics: Annual Medical Costs, All (thousand yen)}
		\begin{tabular}{ccccccccccccc}
			\toprule
			\midrule
			&       &       &       & \multicolumn{9}{c}{Percentiles} \\
			\cmidrule{5-13}    Age   & Obs   & Mean  & Std. Dev. & 1\%   & 5\%   & 10\%  & 25\%  & 50\%  & 75\%  & 90\%  & 95\%  & 99\% \\
			\midrule
			0-4   & 989,193 & 206.0 & 1214.8 & 0   & 5.8   & 15.7  & 44.2  & 93.2  & 174.3 & 320.4 & 502.4 & 1777.9 \\
			5-9   & 968,825 & 104.6 & 388.0 & 0   & 4.3   & 11.0  & 27.9  & 59.0  & 112.9 & 201.1 & 288.5 & 672.9 \\
			10-14 & 962,438 & 76.2  & 379.0 & 0   & 0   & 4.1   & 14.4  & 34.7  & 70.6  & 130.3 & 195.7 & 592.3 \\
			15-19 & 940,088 & 57.6  & 377.4 & 0   & 0   & 0   & 6.3   & 19.8  & 46.1  & 93.5  & 150.8 & 608.1 \\
			20-24 & 967,171 & 51.4  & 348.3 & 0   & 0   & 0   & 2.7   & 14.5  & 38.4  & 86.7  & 151.2 & 600.1 \\
			25-29 & 1,037,876 & 61.8  & 326.7 & 0   & 0   & 0   & 3.8   & 17.4  & 47.0  & 110.7 & 201.6 & 742.5 \\
			30-34 & 1,166,561 & 73.3  & 320.0 & 0   & 0   & 0   & 5.7   & 21.6  & 58.1  & 138.6 & 255.1 & 860.5 \\
			35-39 & 1,310,719 & 81.9  & 378.6 & 0   & 0   & 0   & 6.0   & 23.7  & 65.0  & 157.0 & 282.0 & 949.6 \\
			40-44 & 1,386,228 & 94.0  & 423.2 & 0   & 0   & 0   & 6.1   & 25.7  & 75.2  & 182.0 & 312.3 & 1106.2 \\
			45-49 & 1,183,169 & 119.2 & 489.8 & 0   & 0   & 0   & 7.2   & 32.1  & 99.2  & 228.4 & 381.4 & 1470.9 \\
			50-54 & 964,807 & 159.6 & 622.5 & 0   & 0  & 0   & 10.6  & 48.2  & 138.4 & 291.1 & 479.8 & 2129.6 \\
			55-59 & 765,749 & 206.8 & 744.7 & 0   & 0   & 0   & 16.0  & 72.1  & 177.8 & 354.0 & 597.9 & 2931.1 \\
			60-64 & 581,661 & 262.8 & 894.7 & 0   & 0   & 0  & 26.1  & 101.5 & 218.8 & 427.6 & 771.8 & 3916.0 \\
			65-69 & 221,248 & 334.8 & 936.1 & 0   & 0   & 5.1   & 46.8  & 139.3 & 278.5 & 553.5 & 1094.4 & 4659.9 \\
			70-74 & 101,393 & 485.0 & 1120.0 & 0   & 0   & 21.2  & 98.2  & 216.3 & 415.2 & 884.0 & 1862.1 & 5596.6 \\
			75-79 & 13,292 & 533.5 & 1400.0 & 0   & 0   & 0   & 103.3 & 235.7 & 444.9 & 925.6 & 1878.8 & 6615.0 \\
			\bottomrule
			\bottomrule
		\end{tabular}
		\begin{tablenotes}
			\small
			\item Data Source: The Japan Medical Data Center (JMDC) claim database
		\end{tablenotes}
		\label{tab:table2}
		\end{threeparttable}
\end{ltable}

As for the gender difference, the summary characteristics of medical costs for males and females are presented in Tables \ref{tab:table3} and \ref{tab:table4}.  
Both men and women have a U-shaped medical cost age profile. 
When it comes to children under age 15, the medical cost incurred is, on average, slightly higher for boys.  
However, in the case of middle aged persons, age profiles of medical cost are steeper for females.  
One of the reasons for this gender difference is that the female sample is a mix of self-insured company employees and dependents, as can be seen in Table \ref{tab:table1}.  
It could be that some women have endogenously quit their jobs due to bad health and, unlike their male counterparts, such female individuals might have a greater probability of being included in the sample as dependents.
Thus, we cannot determine whether the difference in comparison with males is of biological nature or due to the sample selection. 
Accordingly, in the empirical analysis we present below, we focus on males' medical costs.
Also, since there are many sample omissions for male individuals over age 60, we restrict our sample to males aged 0-59.

	\begin{ltable}[htbp]
		\centering
		\begin{threeparttable}
		\caption{Summary Statistics: Annual Medical Costs, Males (thousand yen)}
		\begin{tabular}{ccccccccccccc}
			\toprule
			\toprule
			&       &       &       & \multicolumn{9}{c}{Percentiles} \\
			\cmidrule{5-13}    Age   & Obs   & Mean  & Std. Dev. & 1\%   & 5\%   & 10\%  & 25\%  & 50\%  & 75\%  & 90\%  & 95\%  & 99\% \\
			\midrule
			0-4   & 508,679 & 219.7 & 1272.7 & 0   & 6.6   & 17.4  & 47.6  & 99.5  & 185.7 & 344.2 & 542.0 & 1940.1 \\
			5-9   & 497,583 & 113.5 & 434.6 & 0  & 4.9   & 12.1  & 30.1  & 63.8  & 122.6 & 218.6 & 310.4 & 730.8 \\
			10-14 & 495,147 & 83.9  & 417.5 & 0   & 0   & 4.9   & 16.0  & 37.9  & 77.1  & 143.8 & 218.0 & 662.6 \\
			15-19 & 496,162 & 61.2  & 447.0 & 0   & 0   & 0   & 5.6   & 18.9  & 46.1  & 96.3  & 159.0 & 687.3 \\
			20-24 & 568,342 & 47.1  & 386.3 & 0  & 0   & 0   & 0   & 11.3  & 31.1  & 72.6  & 130.6 & 574.3 \\
			25-29 & 633,120 & 50.4  & 351.0 & 0   & 0   & 0  & 1.3   & 12.8  & 34.9  & 82.1  & 150.0 & 589.1 \\
			30-34 & 667,335 & 56.9  & 300.9 & 0   & 0   & 0   & 2.7   & 15.5  & 42.5  & 101.2 & 184.6 & 643.4 \\
			35-39 & 712,794 & 69.3  & 376.6 & 0   & 0   & 0   & 3.6   & 18.2  & 51.8  & 129.1 & 231.2 & 762.8 \\
			40-44 & 751,917 & 87.5  & 423.3 & 0   & 0   & 0   & 4.4   & 21.6  & 68.0  & 172.3 & 291.2 & 999.7 \\
			45-49 & 667,533 & 115.1 & 502.2 & 0   & 0   & 0   & 5.7   & 28.6  & 96.9  & 224.9 & 367.4 & 1379.4 \\
			50-54 & 556,025 & 157.2 & 642.9 & 0   & 0   & 0   & 8.8   & 45.5  & 137.8 & 289.6 & 469.5 & 2087.8 \\
			55-59 & 439,377 & 213.3 & 802.8 & 0   & 0   & 0   & 14.8  & 72.7  & 181.2 & 363.2 & 612.2 & 3030.9 \\
			60-64 & 349,589 & 274.1 & 923.7 & 0   & 0   & 0   & 25.6  & 103.0 & 223.6 & 440.0 & 806.1 & 4171.5 \\
			65-69 & 119,177 & 350.2 & 997.1 & 0   & 0   & 3.6   & 45.1  & 137.4 & 279.8 & 575.9 & 1206.7 & 4966.5 \\
			70-74 & 42,298 & 530.9 & 1278.6 & 0   & 0   & 19.6  & 91.4  & 210.5 & 425.3 & 1025.9 & 2188.0 & 6210.5 \\
			75-79 & 4,822  & 613.4 & 1514.7 & 0   & 0   & 0   & 99.1  & 238.9 & 484.8 & 1160.3 & 2462.4 & 7404.1 \\
			\bottomrule
			\bottomrule
		\end{tabular}
		\begin{tablenotes}
			\small
			\item Data Source: The Japan Medical Data Center (JMDC) claim database
			\end{tablenotes}
		\label{tab:table3}
		\end{threeparttable}
	\end{ltable}

	\begin{ltable}[htbp]
		\centering
		\begin{threeparttable}
		\caption{Summary Statistics: Annual Medical Costs, Females (thousand yen)}
		\begin{tabular}{ccccccccccccc}
			\toprule
			\midrule
			&       &       &       & \multicolumn{9}{c}{Percentiles} \\
			\cmidrule{5-13}    Age   & Obs   & Mean  & Std. Dev. & 1\%   & 5\%   & 10\%  & 25\%  & 50\%  & 75\%  & 90\%  & 95\%  & 99\% \\
			\midrule
			0-4   & 480,514 & 191.4 & 1150.0 & 0   & 5.2   & 14.3  & 41.0  & 87.1  & 162.7 & 295.5 & 458.2 & 1619.3 \\
			5-9   & 471,242 & 95.2  & 331.4 & 0   & 3.5   & 9.9   & 25.8  & 54.4  & 103.3 & 181.4 & 262.2 & 611.6 \\
			10-14 & 467,291 & 68.0  & 333.2 & 0   & 0.   & 3.0   & 13.1  & 31.7  & 64.0  & 116.0 & 171.5 & 511.0 \\
			15-19 & 443,926 & 53.6  & 279.7 & 0  & 0  & 0   & 7.1   & 20.7  & 46.1  & 90.6  & 142.6 & 534.3 \\
			20-24 & 398,829 & 57.5  & 285.5 & 0   & 0   & 0   & 5.7   & 20.3  & 48.9  & 104.3 & 175.3 & 627.2 \\
			25-29 & 404,756 & 79.6  & 283.6 & 0   & 0   & 0   & 8.2   & 27.7  & 67.7  & 152.6 & 285.0 & 890.3 \\
			30-34 & 499,226 & 95.2  & 342.7 & 0   & 0   & 0   & 10.6  & 33.0  & 80.8  & 186.5 & 364.9 & 1031.3 \\
			35-39 & 597,925 & 96.9  & 380.4 & 0   & 0   & 0   & 9.5   & 31.9  & 81.2  & 188.9 & 352.2 & 1102.9 \\
			40-44 & 634,311 & 101.6 & 422.9 & 0  & 0   & 0   & 8.4   & 31.0  & 83.2  & 193.2 & 341.6 & 1194.9 \\
			45-49 & 515,636 & 124.5 & 473.2 & 0  & 0   & 0   & 9.5   & 36.4  & 102.1 & 233.5 & 402.5 & 1566.5 \\
			50-54 & 408,782 & 162.9 & 593.6 & 0   & 0   & 0   & 13.2  & 51.4  & 139.2 & 293.6 & 495.2 & 2167.7 \\
			55-59 & 326,372 & 198.2 & 658.3 & 0   & 0   & 0   & 17.6  & 71.2  & 173.2 & 342.4 & 577.9 & 2748.1 \\
			60-64 & 232,072 & 245.9 & 848.9 & 0   & 0  & 0   & 26.9  & 99.2  & 211.7 & 408.7 & 720.1 & 3522.7 \\
			65-69 & 102,071 & 316.7 & 859.0 & 0   & 0   & 6.5   & 48.8  & 141.4 & 276.9 & 530.7 & 991.3 & 4317.1 \\
			70-74 & 59,095 & 452.1 & 989.8 & 0   & 0   & 23.0  & 102.9 & 220.2 & 410.0 & 807.6 & 1595.1 & 5072.8 \\
			75-79 & 8,470  & 488.0 & 1328.2 & 0   & 0   & 0   & 104.9 & 233.7 & 429.0 & 811.8 & 1512.5 & 5785.7 \\
			\midrule
			\bottomrule
		\end{tabular}
		\begin{tablenotes}
			\small
			\item Data Source: The Japan Medical Data Center (JMDC) claim database
		\end{tablenotes}
		\label{tab:table4}
		\end{threeparttable}
	\end{ltable}
\section{Estimation Method}

We first define five mutually exclusive health transition states for each person-year unit, in accordance with the medical cost in that year.
Previous studies, such as \cite{AKV} and \cite{PP1}, have used quantiles of medical fees for defining individuals' health status.  However, focusing on quantiles makes comparing the health status across different ages difficult.  That is why, in this paper, we define the health status in relation to the level of medical costs incurred by individuals.  
That enables us to take into account medical fee systems such as the above-mentioned high-cost medical care benefit scheme.

We define the health transition states and health statuses as follows (Table \ref{tab:table5}). 
State $Q1$ means that the individual's overall medical cost for that year is between 0 and 7,800 yen, which corresponds to the actual expenditure of 2,340 yen at the co-payment rate of 30\% (which is the usual such rate for adults in Japan, as explained in section 3.1). In the present paper such individuals are regarded as having the best health status, i.e., as being in best health.
Similarly, state $Q2$ means that the annual medical cost is between 7,801 and 24,000 yen (good health), state $Q3$ that it is between 24,001 and 54,000 yen (relatively good health), whereas state $Q4$ denotes the annual cost between 54,001 and 266,999 yen (poor health), and state $Q5$ indicates that the individual's yearly medical cost exceeds 267,000 yen (poorest health). 
These values, except for 267,000 yen, come from the rounded number of medical cost distribution for those who are aged 30-40 and did not pay any medical fees in the previous year: below median for state $Q1$, from the 50th to the 75th percentile for state $Q2$, from the 75th percentile to the 90th percentile for $Q3$ and from the 90th percentile to 266,999 for state $Q4$. 
The value of 267,000 yen, however, is derived from the monthly reimbursement threshold set by the high-cost medical expense benefit scheme. 
Now that we have defined the health transition states based on the distribution of medical costs for those who did not pay any medical fees in the previous year, we turn our attention to the persistency of medical shocks - its magnitude and age-specific differences.

\begin{table}[htbp]
	\centering
	\begin{threeparttable}
	\caption{Definition of Health Status (Japanese yen)}
	\begin{tabular}{ccc}
		\toprule
		\toprule
		State  & Annual Medical Costs & Co-payments (30\%) \\
		\midrule
		$Q1$    & 0 $\sim$ 7,800 & 0 $\sim$ 2,340 \\
		$Q2$    & 7,801 $\sim$ 24,000 & 2,341 $\sim$ 7,200 \\
		$Q3$    & 24,001 $\sim$ 54,000 & 7,201 $\sim$ 16,200 \\
		$Q4$    & 54,001 $\sim$ 266,999 & 16,201 $\sim$ 80,099 \\
		$Q5$    & 267,000 $\sim$ & --  \\
		\bottomrule
		\bottomrule
	\end{tabular}
	\begin{tablenotes}
		\small
		\item Note: In this paper we regard individuals in state Q1 as being in best health condition, those in Q2 as being in good health, whereas Q3 represents relatively good health, Q4 stands for poor health and Q5 for poorest health. Furthermore, it is important to note that we do not observe the actual expenditure when it comes to individuals who have the poorest health status (state $Q5$) because some of them make use of the high-cost medical expenses benefit system and some do not. For states $Q1$ to $Q4$, the co-payment corresponds to the actual expenditure.  
	\end{tablenotes}
	\label{tab:table5}
	\end{threeparttable}
\end{table}

Table \ref{tab:table6} shows fractions of male individuals in each state defined in Table \ref{tab:table5}. 
Except for state $Q5$, which is based on the threshold amount found in the high-cost medical expenses benefit scheme, our state fractions do not show an extreme distribution, suggesting that our definition of health status works well.  
Here we can see that more than a half of middle aged men pay less than 7,200 yen (24,000 $\times$ 0.3 $=$ 7,200 at the co-payment rate of 30\%) a year, thus falling into states $Q1$ and $Q2$, as defined in Table \ref{tab:table5}.
Also, as can be observed in the summary statistics (Table \ref{tab:table3}), after early infancy, the fractions of states $Q4$ and $Q5$ first decrease as individuals grow to become adults, but then start to gradually increase as they get older, thereby creating a U-shaped age profile.
\begin{table}[htbp]
	\centering
	\caption{Fraction of Health Transition States by Age Group, Males}
	\label{tab:table6}
	\begin{threeparttable}
	\begin{tabular}{ccccccc}
		\toprule
		\toprule
		&       & \multicolumn{5}{c}{Fraction of each state ( \% )} \\
		\cmidrule{3-7}    Age   &       & $Q1$    & $Q2$    & $Q3$    & $Q4$    & $Q5$ \\
		\midrule
		0-4   &       & 6.0   & 7.1   & 15.2  & 56.9  & 14.7 \\
		5-9   &       & 7.1   & 12.7  & 23.9  & 49.6  & 6.8 \\
		10-14 &       & 14.3  & 20.9  & 27.6  & 33.6  & 3.6 \\
		15-19 &       & 30.3  & 26.3  & 22.3  & 18.4  & 2.7 \\
		20-24 &       & 41.9  & 26.9  & 17.1  & 11.8  & 2.3 \\
		25-29 &       & 39.3  & 26.6  & 18.0  & 13.5  & 2.6 \\
		30-34 &       & 35.6  & 25.4  & 19.2  & 16.7  & 3.2 \\
		35-39 &       & 33.3  & 23.5  & 19.2  & 20.0  & 4.1 \\
		40-44 &       & 31.4  & 20.9  & 18.0  & 24.1  & 5.6 \\
		45-49 &       & 28.4  & 18.2  & 16.4  & 29.2  & 7.9 \\
		50-54 &       & 23.7  & 14.9  & 14.8  & 35.4  & 11.2 \\
		55-59 &       & 19.0  & 11.9  & 13.0  & 40.8  & 15.3 \\
		\bottomrule
		\bottomrule
	\end{tabular}
	\begin{tablenotes}
		\small
		\item Data Source: The Japan Medical Data Center (JMDC) claim database
		\item We defined the health transition states according to individual's overall medical cost for that year: 0--7,800 yen for $Q1$, 7,801--24,000 yen for $Q2$, 24,001--54,000 yen for $Q3$, 54,001--266,999 yen for $Q4$ and over 267,000 yen for $Q5$.
	\end{tablenotes}
	\end{threeparttable}
\end{table}

\section{Results}
In this section, we describe the four main findings of this paper.
First, we suggest that conditioning only on the state from a previous
year cannot predict a patient's future health expenditure path to the
extent that conditioning on the states from two previous years can. 
Second, we demonstrate that the probability of encountering a
health shock first decreases until around age 10, and then increases with age, particularly, after age 40.
Third, health shock distributions do not differ across age groups until their percentiles reach the median range, but above the median, the health shock distributions of older age groups gradually start to first-order dominate those
of younger groups.
And fourth, we find that the persistency of health shocks also exhibits a
U-shape in relation to age.

\subsection{Markov Chain of Order Two Rather than Order One} 
When Markov chain of order one is used, different types of individuals,
those who experience a certain expenditure level for the first time and
those who have been experiencing that expenditure level for some time,
are mixed together in the same state.  Markov chain of order two is, thus, the
minimal methodological framework that is capable of distinguishing those 
who experience a certain health expenditure level for the first time from those
who have been experiencing that or other levels for some time.
Here, we present several results that indicate the importance of using
Markov chain of order two for investigating the life cycle properties of
the health expenditure process.

\begin{figure}[!hb]
	\centering
	\includegraphics[scale=0.11]{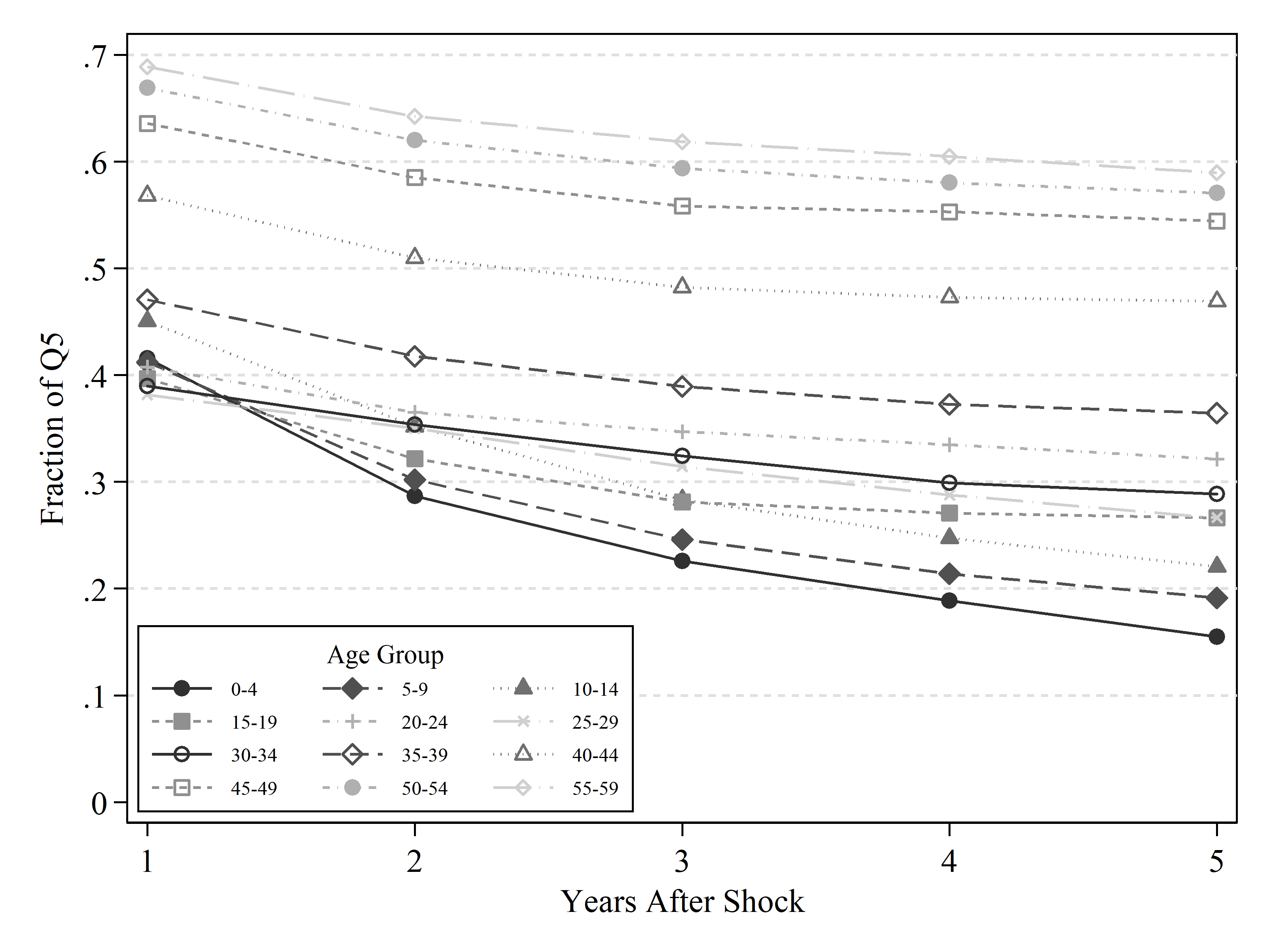}
	\caption{Empirical Frequencies of Transition to State $Q5$ (Poorest Health) from State $Q5$ at $t$} 
	\source{The Japan Medical Data Center (JMDC) claim database}
	\label{k01}
\end{figure}

\begin{figure}[!hb]
	\centering
	\includegraphics[scale=0.11]{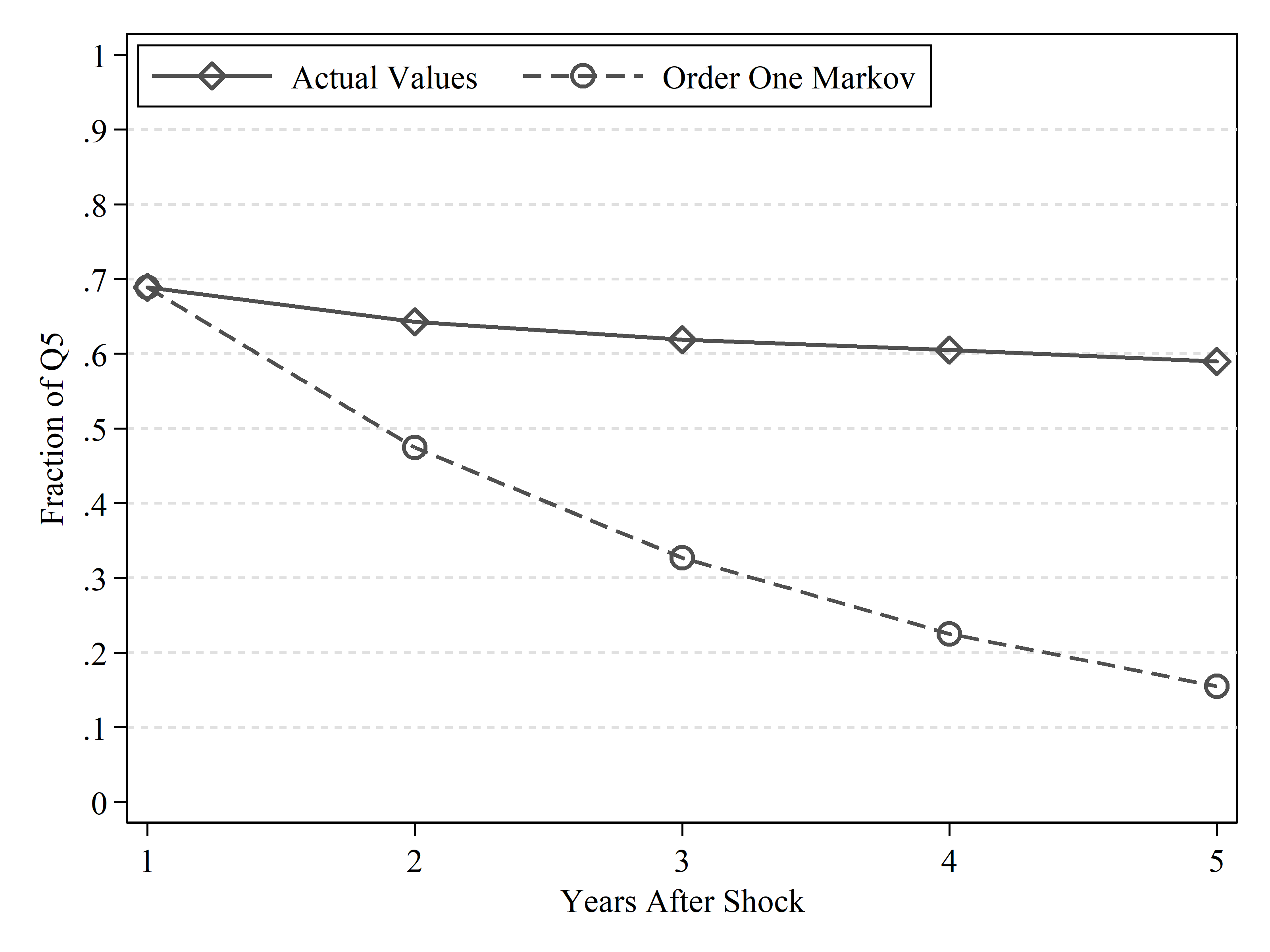}
	\caption{Observed vs Predicted (Order 1 Markov Chain) Frequency of State $Q5$ (Poorest Health) After State $Q5$ in Age Group 55-59}
	\source{The Japan Medical Data Center (JMDC) claim database}
	\label{k17}
\end{figure}

Figure \ref{k01} shows the actual transition frequency to the worst health status (state $Q5$) from the same status ($Q5$) for different age groups.
\footnote{The frequencies are calculated based on the population that remained in the database.  
The omission rate in our data is relatively high. 
For example, among individuals in the age group 55--59, the percentage of those who we could not observe one year later is 8.60\%, and two years later is 44.49\%.
However, we have checked and confirmed that the difference in medical costs between the subjects from each age group who remained in the database and those who dropped out of it is not so large.}
In age group 55--59 (the oldest group), 68.9\% of the subjects who were in state $Q5$
at the year of conditioning maintained that status one year later.  That
means that about 70\% of the people who paid more than 80,100 yen for
health care in a certain year also paid more than 80,100 yen the
following year.  Two years later, the percentage of $Q5$ decreases to
64.2\%.  If the 70\% who were in $Q5$ one year later were the same as those
who had been in state $Q5$ in the year of conditioning, we would expect the
percentage of persons in $Q5$ to decrease to about 49\% ($=$70\%$\times$70\%)
plus a transition to $Q5$ from other states, but since that
possibility is small, we can ignore it in our calculation.  The
actual ratio (64.2\%) is obviously higher than that.  
Figure \ref{k17} shows the same graph for age group 55--59 as in Figure \ref{k01} and 
the hypothetical path of that group if the frequency of those who remained in the same health state one year later (68.9\%) continued in the following years.

As can be observed from the figure, the rate of decrease in the percentage of subjects with the worst health status (state $Q5$) is very
slow - even five years later 59.0\% of the subjects remain in state $Q5$.  This result
suggests that, among the people in state $Q5$, there are two main types of
patients - the high expenditure on health for some of them is due to
``temporary'' health shocks, and such subjects have enough chance to
restore their health, but on the other hand, some patients' large
spending is due to more ``continuous'' health shocks, so many of them  
will have to continue paying large sums at present and in the future.
To illustrate this more point clearly, next we show how future health
expenditure paths differ depending on the medical cost from the
previous two years.

\begin{figure}[!ht]
	\centering
	\includegraphics[scale=0.11]{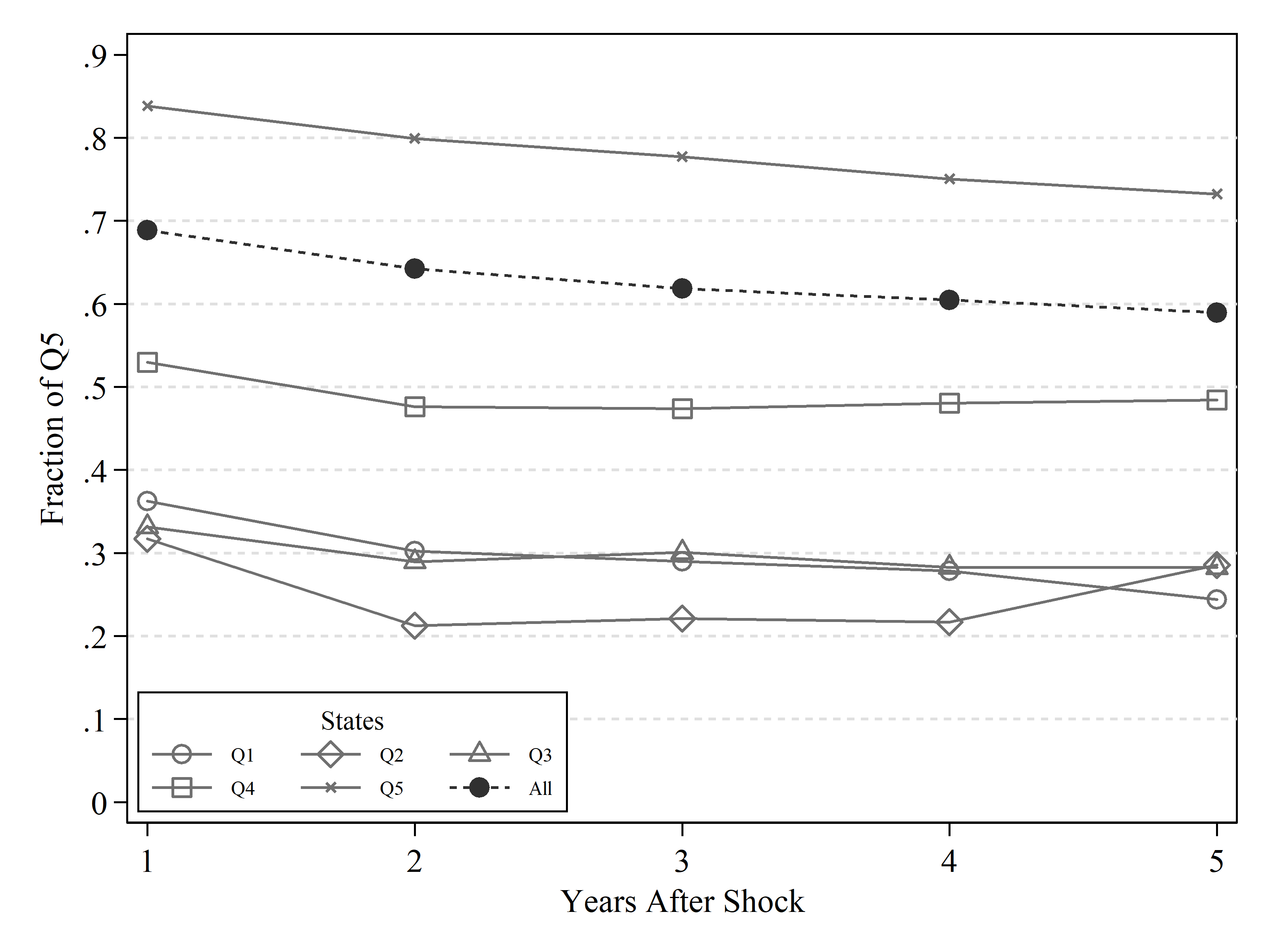}
	\caption{Empirical Frequencies of State $Q5$ (Poorest Health) from State $Q5$ at $t$ with
		Different States at $t-1$ for Age Group 55--59} 
	\source{The Japan Medical Data Center (JMDC) claim database}
	\label{k02}
\end{figure}

Figure \ref{k02} indicates the empirical transition probabilities for the age 
group 55--59, but in it the sample for that age group is divided in
accordance with the subjects' health expenditure from two years
before the year of conditioning.  The graph ``all'' in the figure is
identical with the one in Figure \ref{k01}.  The graphs in Figure \ref{k02} clearly
indicate that persons in the same state $Q5$ have dramatically different
probabilities to remain in it in the following years, depending on their
state from two years before.  Thus, if individuals in the age group
under discussion here paid over 80,100 yen for their health (i.e., if they were in state $Q5$) in a certain year, but their expenditure was under 16,200 yen in
the previous year (i.e., they were in state $Q1$, $Q2$ or $Q3$ in the previous year), then only about
one-third of them will remain in state $Q5$ in the following year.  However, if
their medical cost was over 267,000 yen (in state $Q5$) in the previous year too, then more than 80\% (83.9\%) of them will remain in state $Q5$ the
following year.  Similar results are observed for all other age groups,
too.

This clear difference in health expenditure paths indicates that the health status of the individuals who were in
the same $Q5$ state is varied - some were temporarily in poor health, with
many among them cured the following year, while others seemed to be
in poor health for a longer period, with most of them remaining in state $Q5$
for a protracted period of time.  Thus, by taking into account the subjects' medical costs for the previous year in addition to the current year, we can, at least partially, decompose mixed populations into their ``real'' health statuses.

\subsection{Initial Health Shock Occurrence Probabilities and their Distribution} 
In this section, we examine how initial health shock probabilities differ across
age groups.  We also examine how the distribution of the magnitude of the health shocks differs
across age groups.

\begin{figure}[!hb]
	\centering
	\includegraphics[scale=0.11]{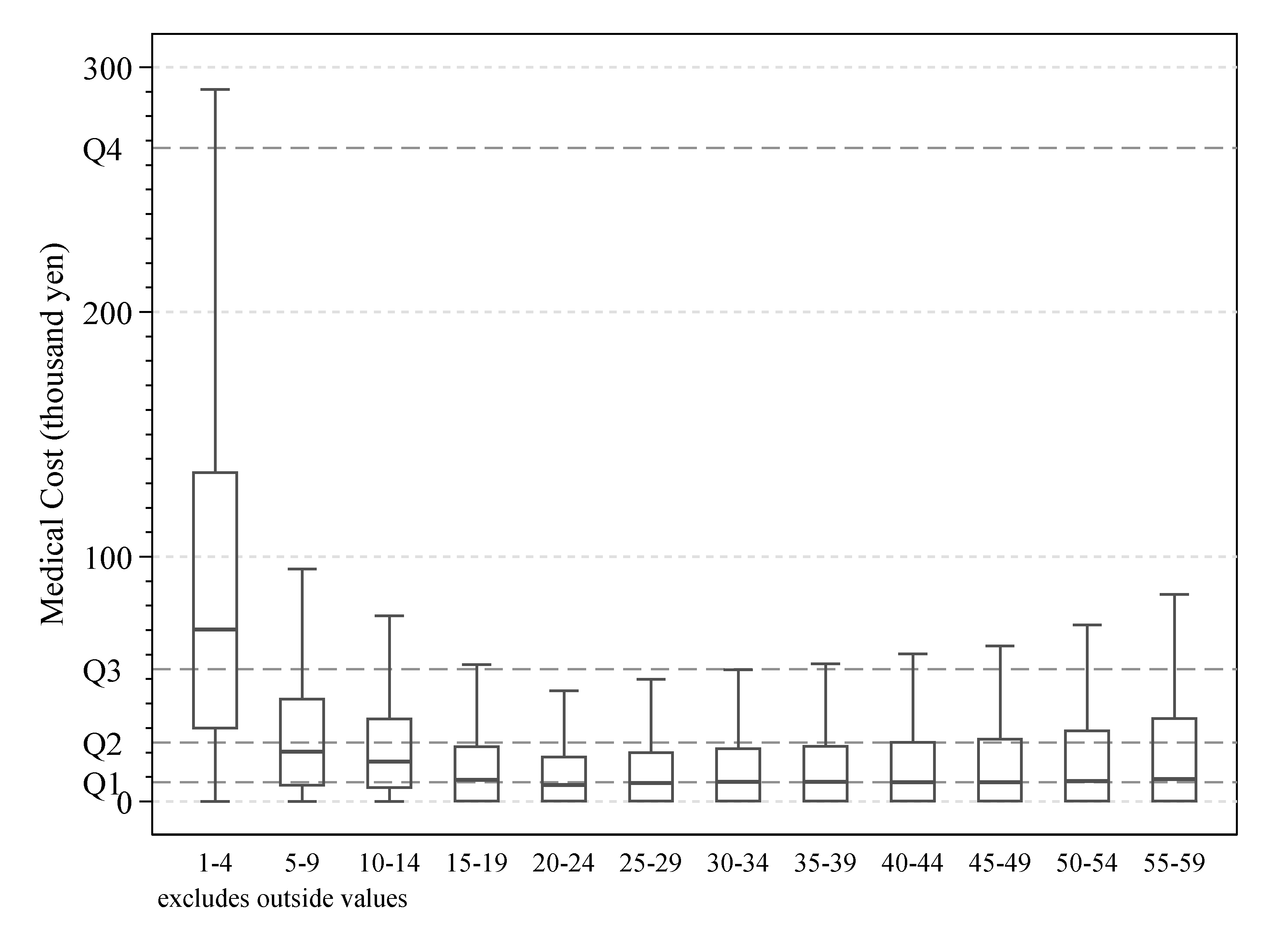}
	\caption{Box Plots of Medical Costs at Age $t$ Given that the Subjects were in Best Health (state $Q1$) at Age $t-1$ (Male)} 
	\source{The Japan Medical Data Center (JMDC) claim database}
	\label{k03}
\end{figure}

\begin{figure}[!ht]
	\centering
	\includegraphics[scale=0.11]{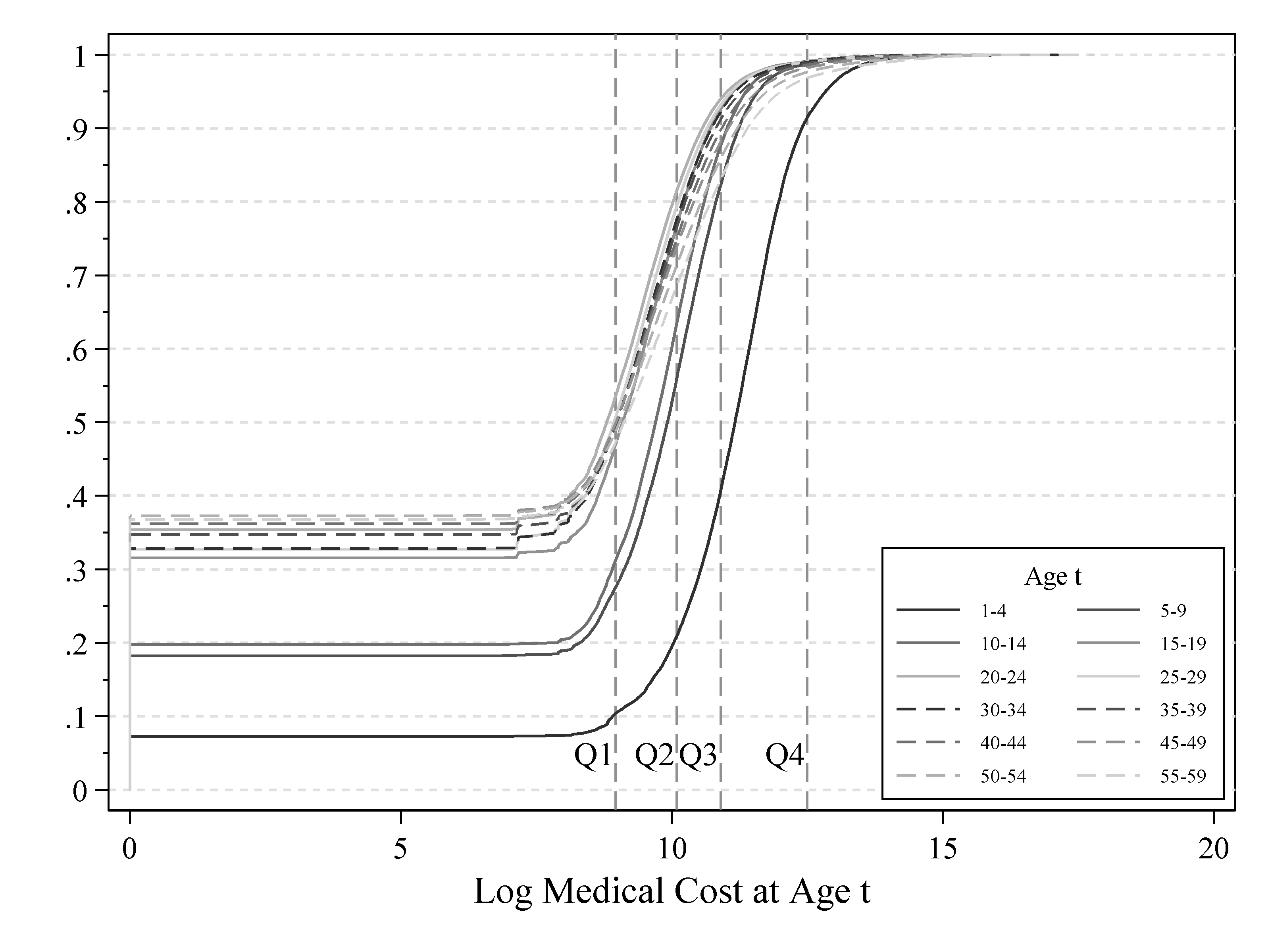}
	\caption{Empirical CDFs of the Logarithm of Medical Costs at Age $t$ Given that the Subjects were in Best Health (state $Q1$) at Age $t-1$ (Male)} 
	\source{The Japan Medical Data Center (JMDC) claim database}
	\label{k04}
\end{figure}

Figure \ref{k03} indicates box plots of the annual medical costs for each
age group, conditioning on that they are in the best health condition (state $Q1$) in the previous year.
This practically means that we take up people who experienced a health problem
after having spent virtually no money on health the year before.  This
justifies calling the expenditures ``health shocks.''  Figure \ref{k04}
shows the distribution of ``health shock'' magnitude by age groups.

These figures suggest that, except for age groups under 15, the
distribution of health shocks is roughly similar for all age groups up
to the group 40--45.  After that age group, a small increasing trend in
accordance with the rise in age is visible, especially for health
shocks beyond the median level.  In particular, the graphical expositions in the two figures, along with the
dotted lines which indicate the threshold values that define the five
states, seem to show that each health state defined by these values contains a
certain number of people from all age groups, except from the group age 0 to 5.  If that is the case, defining
the states based on absolute expenditure, and not percentiles, should allow
us to interpret the meaning of each state more easily and to standardize
health shocks for different age groups.\\


\begin{ltable}[htbp]
	\centering
	\begin{threeparttable}
	\caption{ Annual Medical Fees by Transition Path (thousand yen)}
	\begin{tabular}{ccccccccccccccc}
		\toprule
		\toprule
		&       & \multicolumn{6}{c}{$Q1$ $\rightarrow$  $Q5$}                &       & \multicolumn{6}{c}{$Q5$ $\rightarrow$  $Q5$} \\
		\cmidrule{3-8}\cmidrule{10-15}    Age   &       & Obs   & Mean  & Std. Dev. & Median & Min   & Max   &       & Obs   & Mean  & Std. Dev. & Median & Min   & Max \\
		\midrule
		0-4   &       & 2152  & 622.5 & 1233.5 & 409.6 & 267.0 & 26859.7 &       & 24123 & 861.8 & 2429.6 & 425.9 & 267.0 & 110920.5 \\
		5-9   &       & 279   & 772.2 & 1586.2 & 452.8 & 267.2 & 20241.5 &       & 10888 & 856.4 & 1910.3 & 391.8 & 267.0 & 45312.3 \\
		10-14 &       & 525   & 794.1 & 1187.4 & 482.5 & 267.1 & 16766.2 &       & 6034  & 1536.3 & 3164.6 & 522.9 & 267.0 & 84139.3 \\
		15-19 &       & 1142  & 987.5 & 1576.8 & 563.7 & 267.7 & 18180.7 &       & 3545  & 1588.5 & 3617.4 & 563.3 & 267.1 & 54111.8 \\
		20-24 &       & 1728  & 831.4 & 986.1 & 518.3 & 267.1 & 12339.4 &       & 3525  & 1498.1 & 3473.0 & 585.3 & 267.0 & 112965.0 \\
		25-29 &       & 1849  & 851.9 & 1280.7 & 502.1 & 267.0 & 18791.0 &       & 5041  & 1352.6 & 2710.8 & 532.0 & 267.1 & 55967.3 \\
		30-34 &       & 1977  & 852.1 & 1206.2 & 501.0 & 267.2 & 18408.1 &       & 7375  & 1099.0 & 1948.5 & 493.1 & 267.0 & 63656.5 \\
		35-39 &       & 2304  & 904.3 & 1345.1 & 503.4 & 267.3 & 26664.1 &       & 11806 & 1025.7 & 2041.2 & 471.9 & 267.0 & 51613.6 \\
		40-44 &       & 2779  & 967.2 & 1349.6 & 537.6 & 267.0 & 22202.5 &       & 18342 & 997.6 & 1859.6 & 467.4 & 267.0 & 44747.9 \\
		45-49 &       & 2856  & 1102.6 & 1521.5 & 550.4 & 267.0 & 23297.5 &       & 24825 & 975.5 & 1814.7 & 458.3 & 267.0 & 43114.6 \\
		50-54 &       & 2587  & 1212.3 & 1630.1 & 619.7 & 267.1 & 21576.5 &       & 29589 & 996.7 & 1887.8 & 460.6 & 267.0 & 57328.6 \\
		55-59 &       & 2285  & 1361.2 & 2102.3 & 646.4 & 267.0 & 53521.2 &       & 32955 & 1092.8 & 2067.2 & 458.6 & 267.0 & 51045.1 \\
		\bottomrule
		\bottomrule
	\end{tabular}
	\begin{tablenotes}
		\small
		\item Data Sources: The Japan Medical Data Center (JMDC) claim database 
		\item We defined the health transition states according to individual's overall medical cost for that year: 0--7,800 yen for $Q1$, 7,801--24,000 yen for $Q2$, 24,001--54,000 yen for $Q3$, 54,001--266,999 yen for $Q4$ and over 267,000 yen for $Q5$.
	\end{tablenotes}
	\label{tab:table7}
	\end{threeparttable}
\end{ltable}

In order to make the differences in medical costs between patients who are in the same state but have transitioned from other states in the previous year visible,
we present Table 7, which shows the summary statistics of medical costs of patients in the poorest health (state $Q5$) divided by the age group and the state in the previous year (Here we only show the transition paths from the states $Q1$ and $Q5$).
Roughly speaking, the median values for each age group in these two transition paths do not differ dramatically.
On the other hand, mean values show a more obvious difference, especially for those in their teens and twenties, but this is mainly due to the very large difference in the maximum values.
Therefore, we can say that, for most individuals in the same current state, the medical cost is not so different in relation to their state in the previous year.

\begin{figure}[!hb]
	\centering
	\includegraphics[scale=0.11]{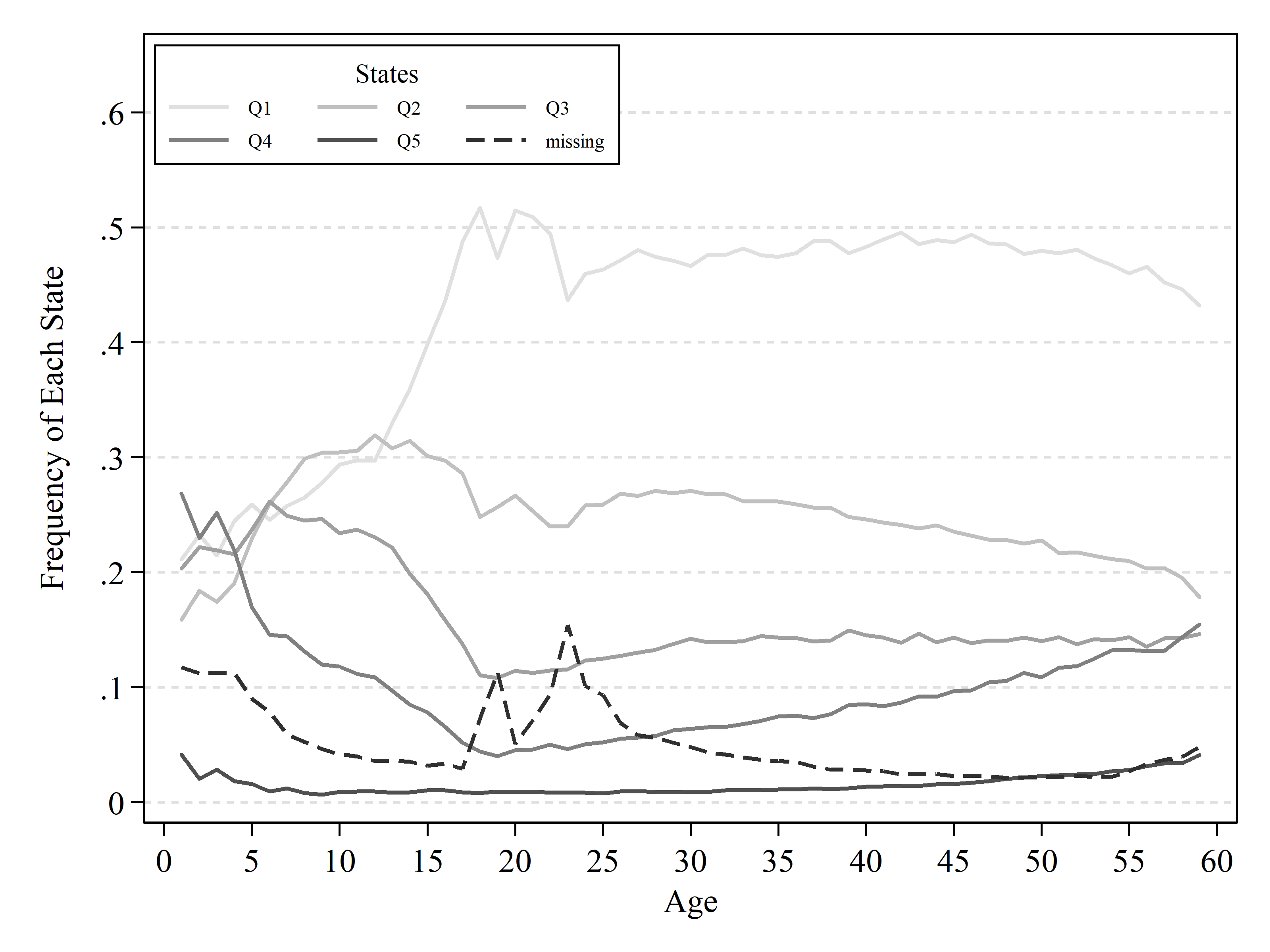}
	\caption{Empirical Frequency of Suffering a Health Shock at Age $t$ Given that the Subjects Were in Best Health (State $Q1$) at Age $t-1$ (Male)} 
	\source{The Japan Medical Data Center (JMDC) claim database}
	\label{k05}
\end{figure}

Next, in Figure \ref{k05}, we show the frequency of each state for each age, including a ``missing'' state, given
that the subjects were in state $Q1$ the previous year. This figure can be interpreted as the change in the
frequency of health shocks for healthy people in relation to their age.

\begin{figure}[!ht]
	\centering
	\includegraphics[scale=0.11]{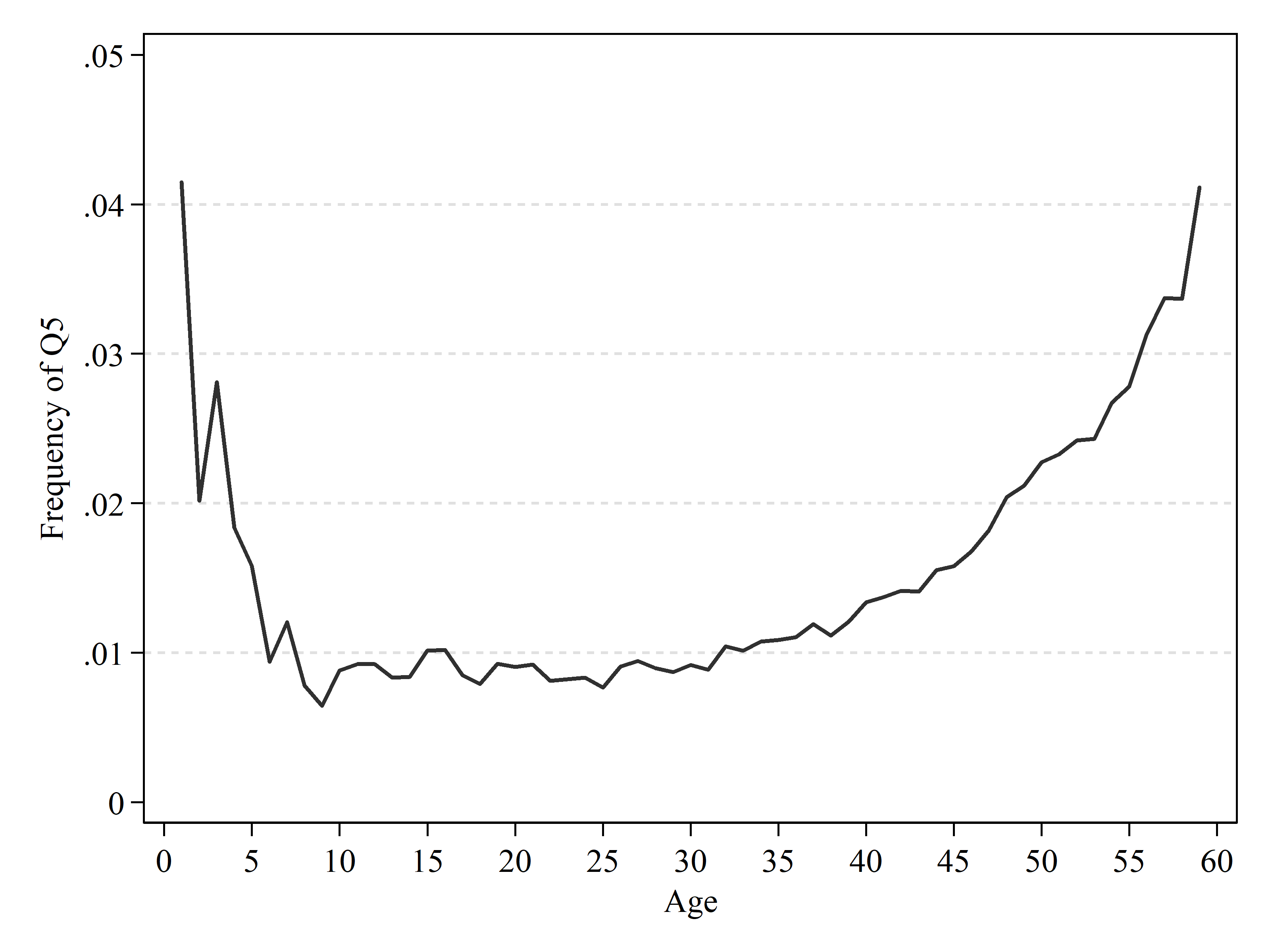}
	\caption{Empirical Frequency of Suffering a Large Health Shock ($Q5$) at Age $t$ Given that the Subjects Were in Best Health (state $Q1$) at Age $t-1$ (Male)} 
	\source{The Japan Medical Data Center (JMDC) claim database}
	\label{k06}
\end{figure}

We can see that the frequency of state $Q5$ is very low for all ages (in fact,
it is below 4 \% for all ages except age 1 and the ages over 59), but
that it increases slightly with age.  To view this in more detail, see  
Figure \ref{k06} in which the frequency of state $Q5$, extracted from Figure \ref{k05}, has
been plotted.

From age 1 to 9, the probability of experiencing a major health shock decreases, recording the minimum value
of 0.64\% at age 9.  From age 10, however, it gradually increases with
age, reaching 4.11\% at the age of 60.\footnote{Since the database we utilize consists of information on company
employees, a large selection issue presents itself for ages beyond 60, since that was the age of mandatory retirement in the period observed.  For this reason, we
choose not to report results for ages 61 and above.}
In particular, the slope of the graph becomes steeper over age 40.  The
percentage is still low around age 60, but health shocks occur
approximately four times more often to people who are 60 years old than
to those who are in their 20s or 30s (around 1\%).  This is not a small
difference when we consider people's dynamic behaviors, e.g. asset
accumulation or participation in health insurance.

\begin{figure}[!hb]
	\centering
	\includegraphics[scale=0.11]{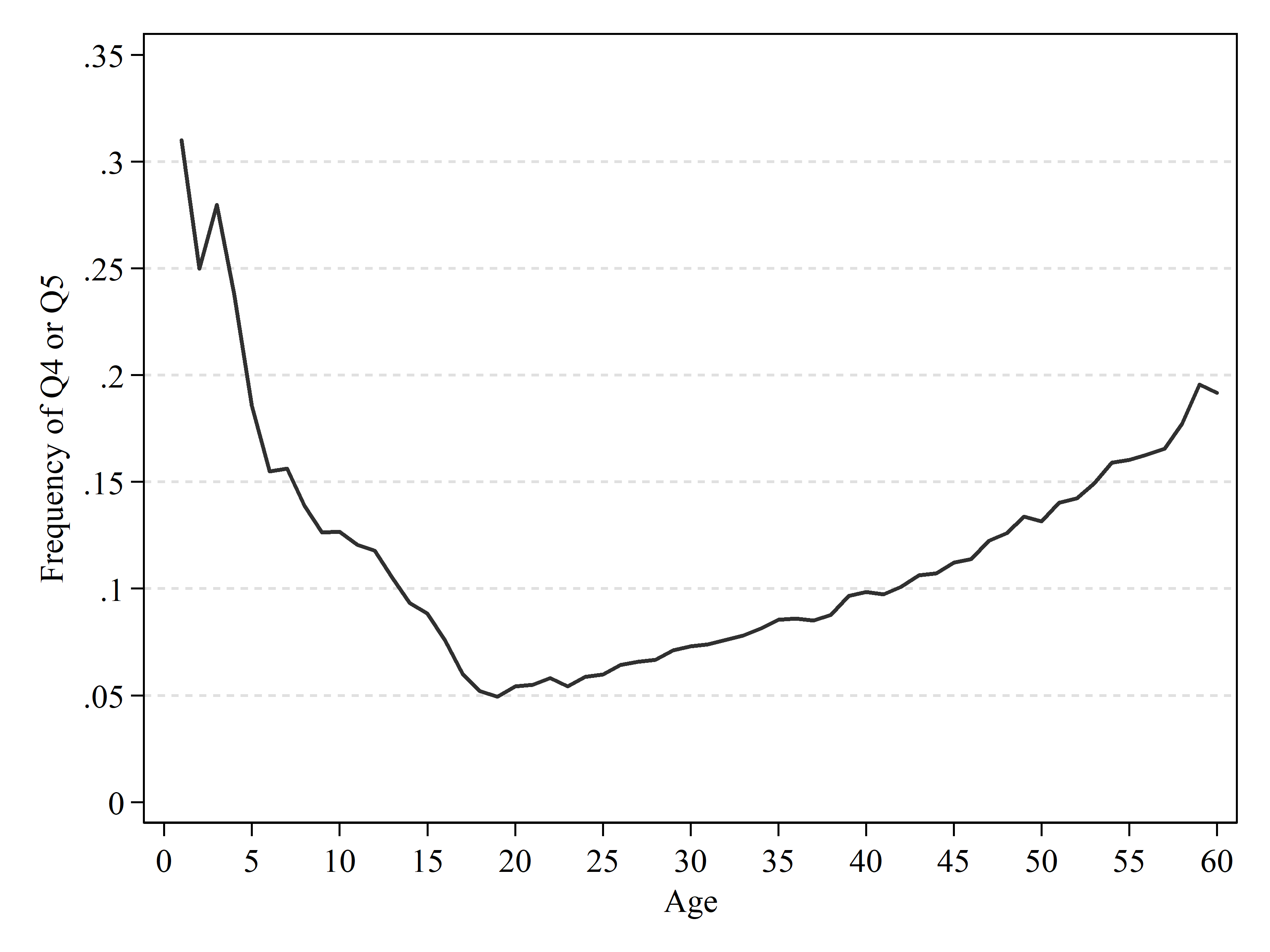}
	\caption{Empirical Frequency of Suffering a Large Health
		Shock ($Q4$ or $Q5$) at Age $t$ Given that the Subjects Were in Best Health (state $Q1$) at Age $t-1$ (Male)} 
	\source{The Japan Medical Data Center (JMDC) claim database}
	\label{k07}
\end{figure}

It is worth noting that the presented result is not due to the setting of the threshold
level. Figure \ref{k07} shows a graph similar to that in Figure \ref{k06}, but
indicates the frequency of states $Q4$ and $Q5$, i.e., the percentage
of people who pay over 16,200 yen for health in a given year but paid only
2,340 yen or less in the previous year. Here too we can find a similar
increasing trend.

\begin{figure}[!ht]
	\centering
	\includegraphics[scale=0.11]{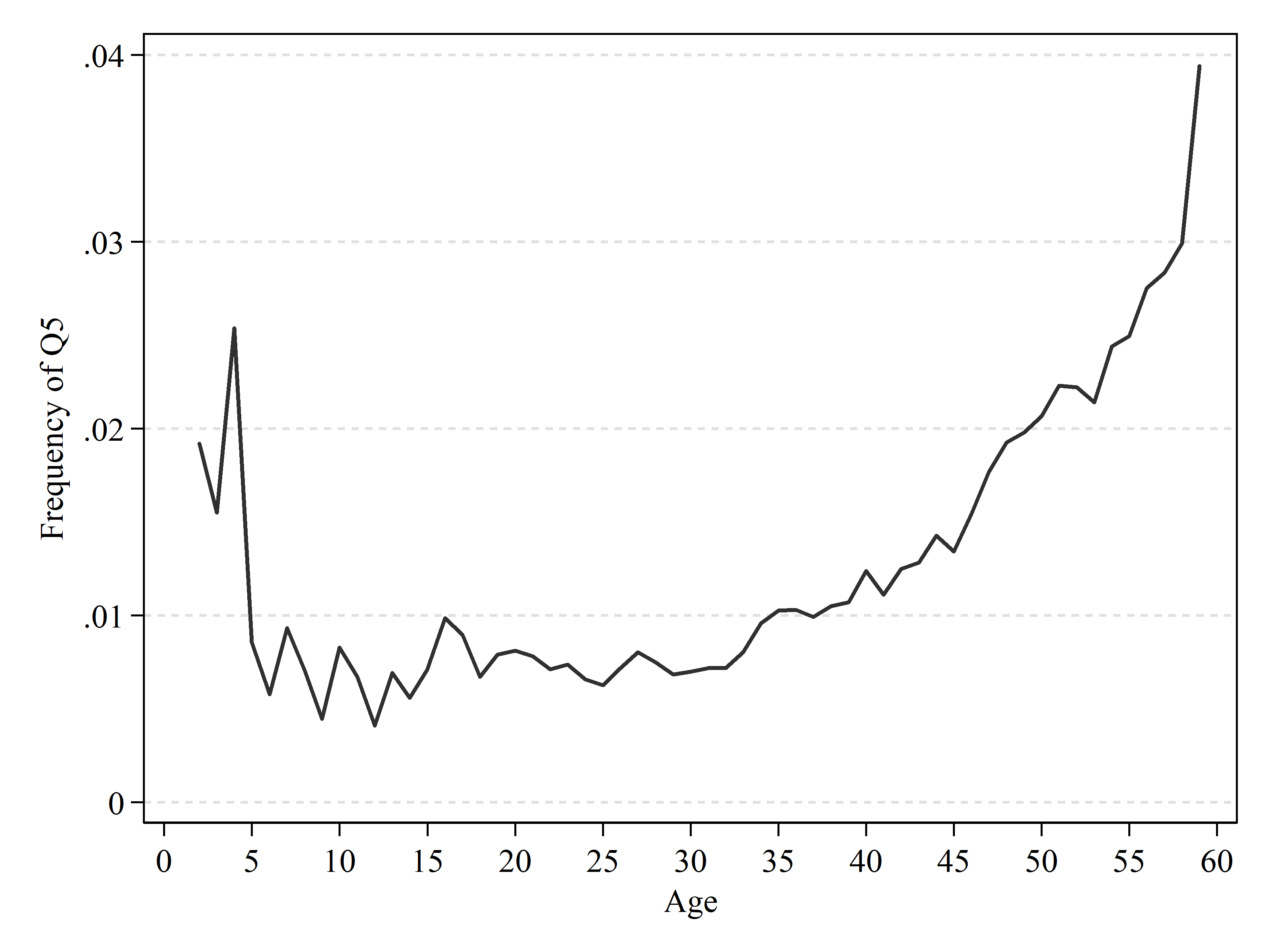}
	\caption{Empirical Frequency of Poorest Health $Q5$ at Age $t$ Given that the Subjects Were in Best Health (state $Q1$) Both at Age $t-1$ and Age $t-2$ (Male)} 
	\source{The Japan Medical Data Center (JMDC) claim database}
	\label{k08}
\end{figure}

We also observe similar results when we condition the sample on the best health status (state $Q1$) both at age $t-1$ and $t-2$ (Figure \ref{k08}).


To sum up, these figures suggest that the probability of a health shock
increases as age increases, especially after age 40.  This result
indicates that when we are constructing a consumer's dynamic optimization model
with medical costs, we need to model the transition probabilities
of health status or health expenditure age-dependent if we want to
incorporate the heterogeneous behaviors of consumers as they age.

\subsection{Persistency After Health Shocks}
\begin{figure}[!hb]
	\centering
	\includegraphics[scale=0.11]{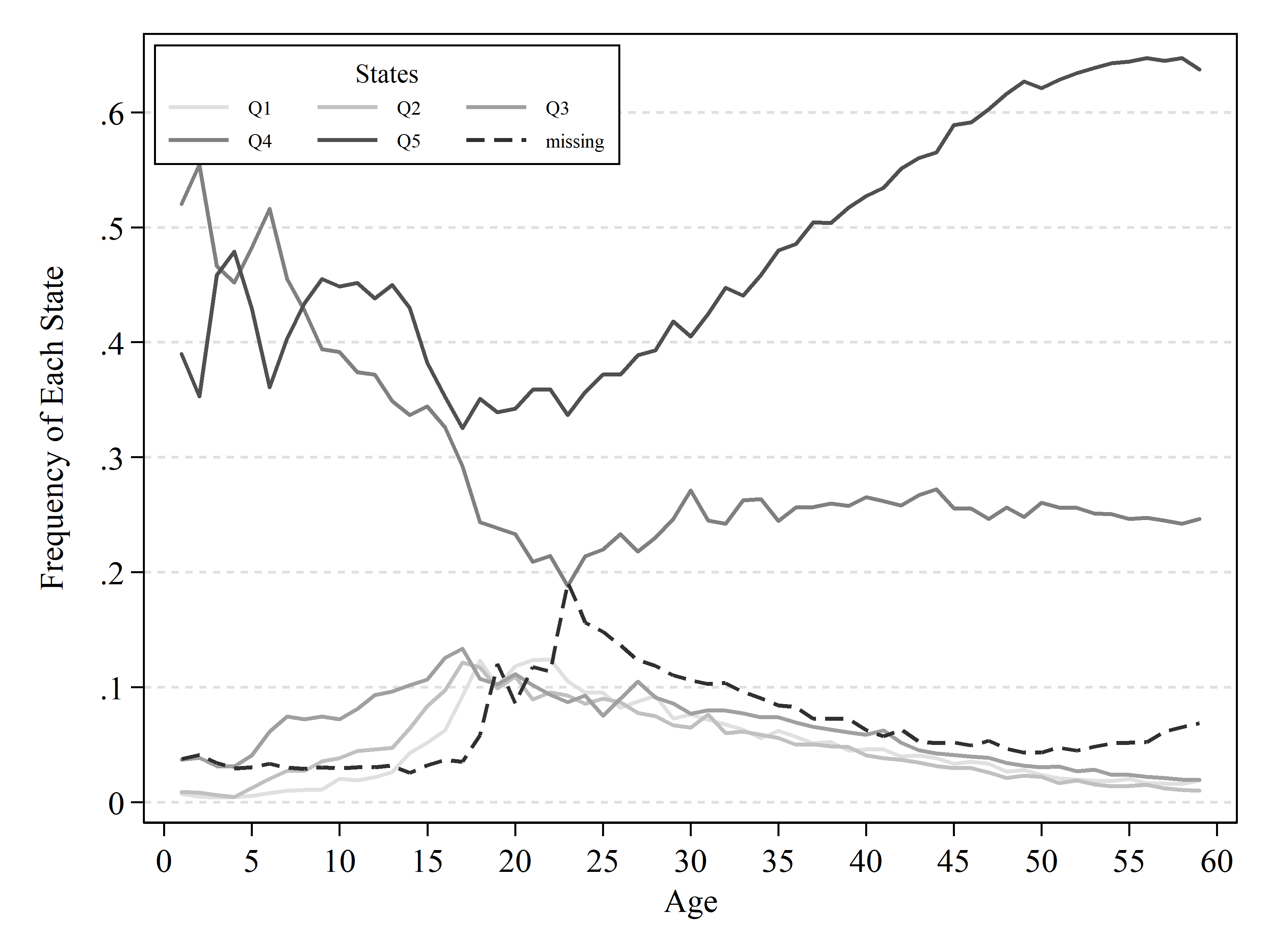}
	\caption{Empirical Frequency of Poorest Health (state $Q5$) at Age $t$ Given that the Subjects Were in Poorest Health ($Q5$) at Age $t-1$ (Male)} 
	\source{The Japan Medical Data Center (JMDC) claim database}
	\label{k09}
\end{figure}

In this section, we will see that the ``persistency'' of health shocks
differs according to age.  First, in Figure \ref{k09}, we indicate the empirical  
transition probabilities to each state from state $Q5$ for each age group.

Compared to Figure \ref{k05}, the probability of high expenditure (states $Q4$ or $Q5$)
is clearly higher, which suggests that once people suffer a health
shock, they tend to stay in poor health the following year, meaning that
the adverse effect of health shocks on health (and on health
expenditure) persists over years.

Furthermore, the persistency differs across ages.  We can see that the
frequency of state $Q5$, the highest health expenditure group, fluctuates
between about 30\% and 45\% for the young ages (ages 1 to 25), and
then starts to increase with age.  For ages over 50, the
probability of state $Q5$ is over 60\% --- roughly double the value for ages
around 25.  We can, therefore, say that health shock persistency 
is higher for the very young and for those over 45.

\begin{figure}[!ht]
	\centering
	\includegraphics[scale=0.11]{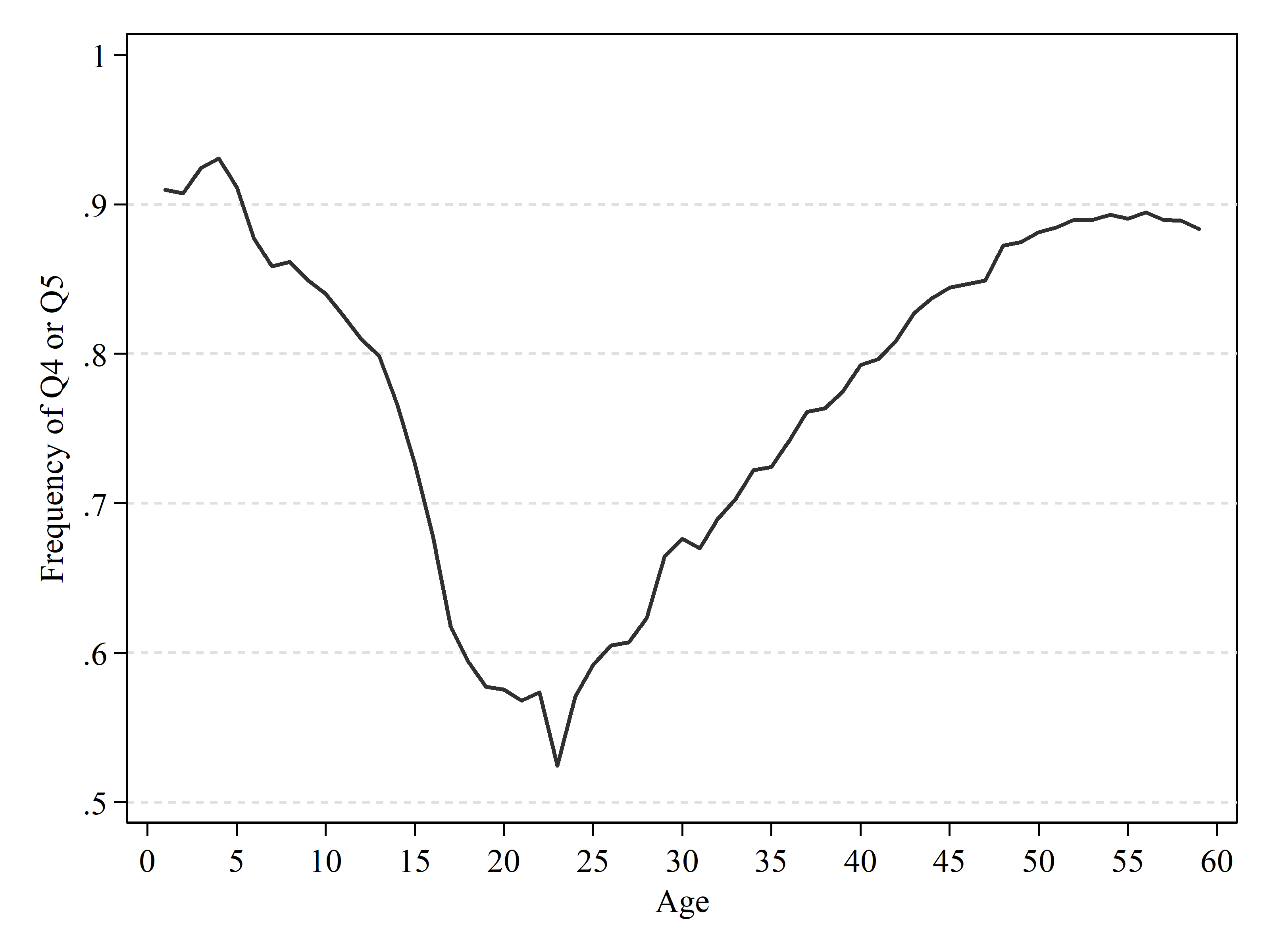}
	\caption{Empirical Probabilities of Transitions from State $Q5$ to States $Q4$ or
		$Q5$ at Age $t$ (Males)} 
	\source{The Japan Medical Data Center (JMDC) claim database}
	\label{k10}
	
\end{figure}

Similar to Figure \ref{k07}, here too we can see that the persistency is not
due to the level of the threshold set.  Figure \ref{k10} displays a graph
similar to that in Figure \ref{k09} but indicates the empirical transition
probabilities to states $Q4$ or $Q5$ from $Q5$, e.g. the percentage of people who pay
over 16,200 yen per year to maintain their health.  Although the
persistency for children, in particular those below 10, becomes high and
shows almost the same magnitude as in those over 45, the results are
similar to the results in Figure \ref{k09}. 

\begin{figure}[!hb]
	\centering
	\includegraphics[scale=0.11]{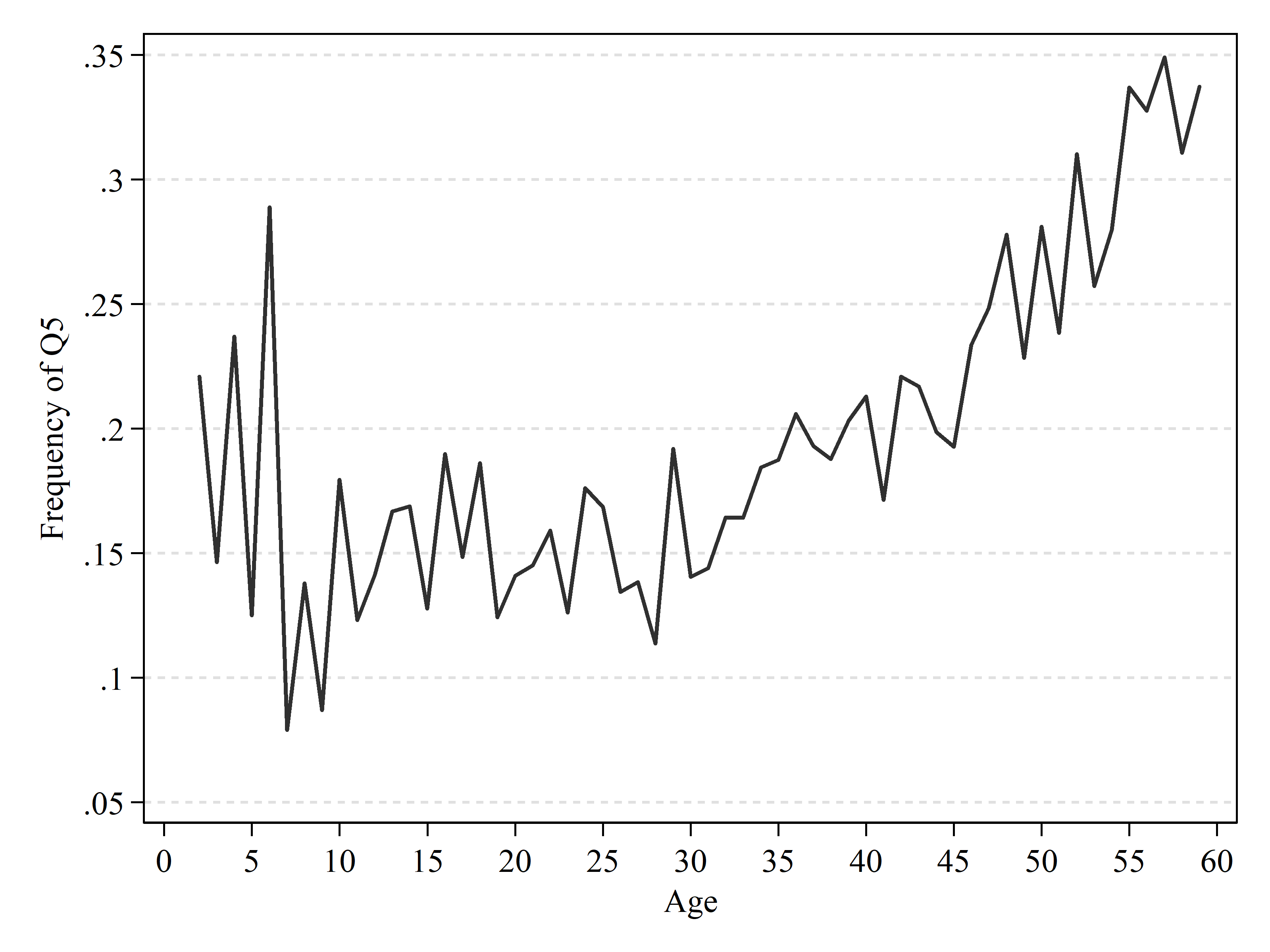}
	\caption{Empirical Probability of Transition to State $Q5$ at Age $t$
		Given that the Subjects Were in State $Q5$ at Age $t-1$ and State $Q1$ at Age $t-2$ (Males)} 
	\source{The Japan Medical Data Center (JMDC) claim database}
	\label{k11}
\end{figure}

Considering our previous argument that in the state $Q5$ there are both people
with temporary and with relatively continuous health issues, the  
higher persistency for older people may simply reflect the fact that a larger
portion of them is in state $Q5$ with continuous health problems, which accumulate with age.   
To help assess this possibility, in Figure \ref{k11}
the same frequencies of state $Q5$ are plotted, but this time the conditioning is not only on $Q5$ for the
year before, but also on state $Q1$ for two years before.  Thus, it may be said that this graph
focuses on individuals who have experienced a large health shock.

Due to the small sample size (there are only around 50 subjects of each
age under 10), the graph starts to fluctuate, in comparison to the
corresponding graph in Figure \ref{k09}.  Nonetheless, we can also observe
that the persistencies become higher as age increases.  Thus, this graph
too reinforces our finding regarding persistency and age.

Note, however, how the magnitude of the transition probabilities drops
to about 35\% in Figure \ref{k09} from over 60\%  in Figure \ref{k11} for those who are 60.  This again shows  
the importance of using at least the order two Markov chain model.

\begin{figure}[!ht]
	\centering
	\includegraphics[scale=0.11]{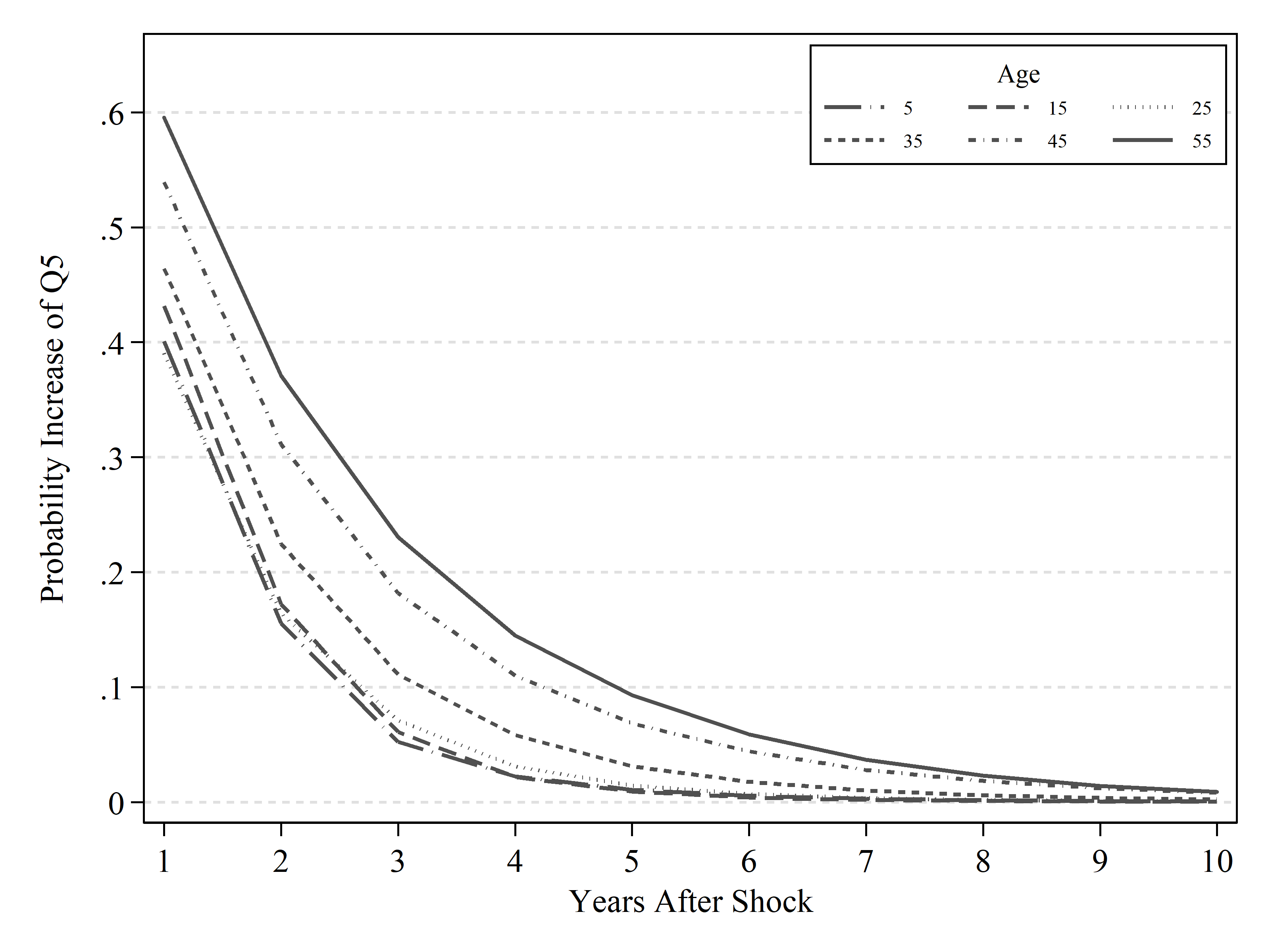}
	\caption{Differences in Empirical Probabilities of Transition
		to Poorest Health (state $Q5$) from Best Health (state $Q1$) and from Poorest Health (state $Q5$)
		(Males)} 
	\source{The Japan Medical Data Center (JMDC) claim database}
	\label{k12}
\end{figure}

To see how persistency varies across ages, we conduct another
analysis.  First, we calculate the transition probabilities to each
state, conditional on the state one year before for each age.  
Then, by designating one state at some age, we calculate
transition probabilities for each state at each age for the future using the estimated probabilities.  
Finally, we compare the difference between the calculated transition probabilities which
start from the best health status (state $Q1$) with those which start from
the worst health status (state $Q5$), and interpret these differences as the
effects of a health shock over time.  Since the effect of the first
state vanishes as we iterate the probabilities repeatedly, the
difference goes to zero as years pass.  Figure \ref{k12} shows these
differences for several start ages.

\begin{figure}[!hb]
	\centering
	\includegraphics[scale=0.11]{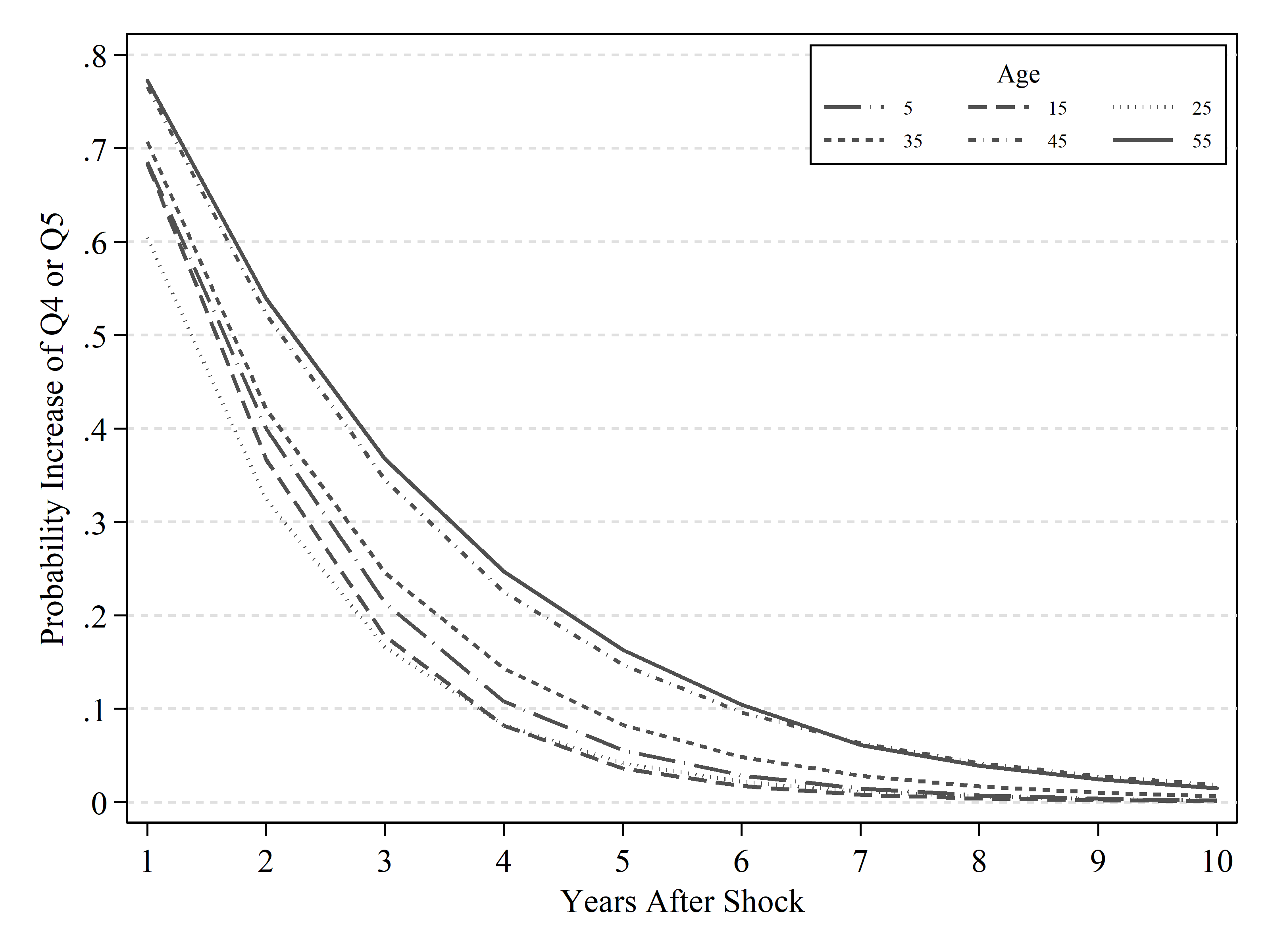}
	\caption{Differences in Empirical Frequencies of
		Transition to Poorest and Poor Health (states $Q4$ or $Q5$) at Age $t$ from Best Health (state $Q1$) vs from Poorest Health (state $Q5$) at Age $t-1$ (Males)} 
	\source{The Japan Medical Data Center (JMDC) claim database}
	\label{k13}
\end{figure}

It is clear that the probability increase of state $Q5$ is the highest for all
the following years when one starts from the worst health status at
age 55, as well as that it is second highest when it starts from age 45. Although the
probability differences almost vanish in six years (becoming less than 1\%) for
relatively younger ages (under 35), the difference that starts from age
55 is 6.20\% and that from age 45 is 4.50\%. Therefore, we can confirm
that health shock persistency becomes higher and longer as age
increases.  When we check the probability increase of states $Q4$ and $Q5$ combined, and not only
of state $Q5$, we come to similar results. (Figure \ref{k13}).

\begin{figure}[!ht]
	\centering
	\includegraphics[scale=0.11]{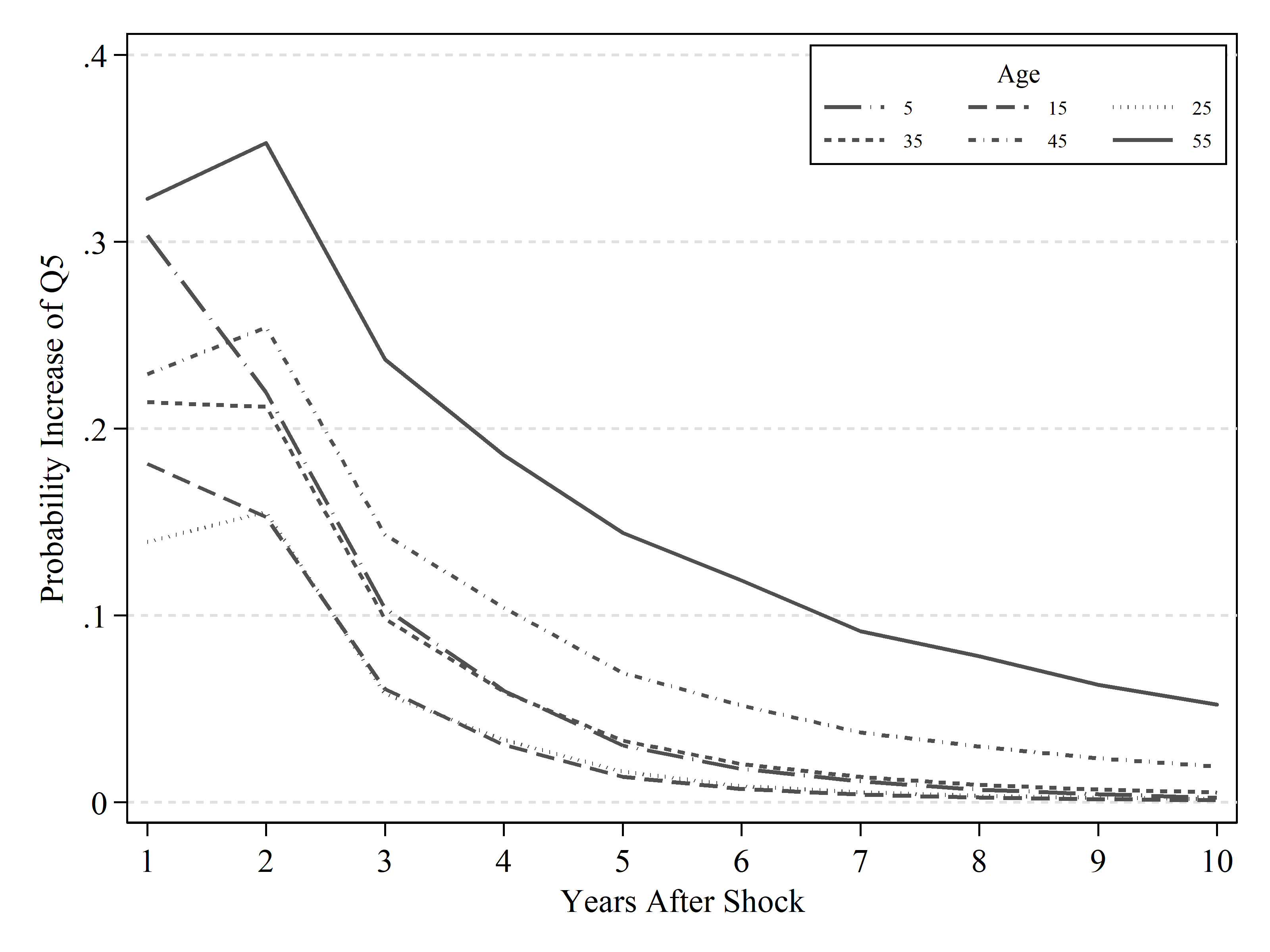}
	\caption{Differences in Empirical Frequencies of Transition to Poorest Health (state $Q5$) at Age $t$ from Best Health (state $Q1$) at both Ages $t-1$ and $t-2$ vs from Best Health ($Q1$) at Age $t-2$ and Poorest Health ($Q5$) at Age $t-1$ (Male)} 
	\source{The Japan Medical Data Center (JMDC) claim database}
	\label{k14}
\end{figure}

Again, having in mind the existence of the mixture of types of people in the worst health status (state $Q5$), we conduct the same
analysis, this time conditioning on two states.  We calculate the transition
probabilities to each state, conditional on the state one year before
and two years before (therefore, using $5\times5=25$ patterns) for each age,
and then plot the differences between the calculated transition
probabilities for transitions that start from the best health status (state $Q1$) both one and two years before, and those that
start from state $Q5$ one year before and from state $Q1$ two years before.  We see a
similar relationship in Figure \ref{k14}.  However, again, it is worth noting that the
level of persistency changes drastically and that those who are 35 years old now show more
persistency than those who are 15 or 25, and become similar to children age 5.

In this section, we have focused on the analysis of five health transition states defined in relation to the medical costs incurred.  This is because, as we have already
explained, medical costs are highly skewed, which is why
regression results using exact amounts of incurred cost may be distorted
by outliers.  Nonetheless, to demonstrate the robustness of our results
using the five states, here we present some simple regression results
obtained by using exact amounts of medical cost.

\begin{figure}[!hb]
	\centering
	\includegraphics[scale=0.11]{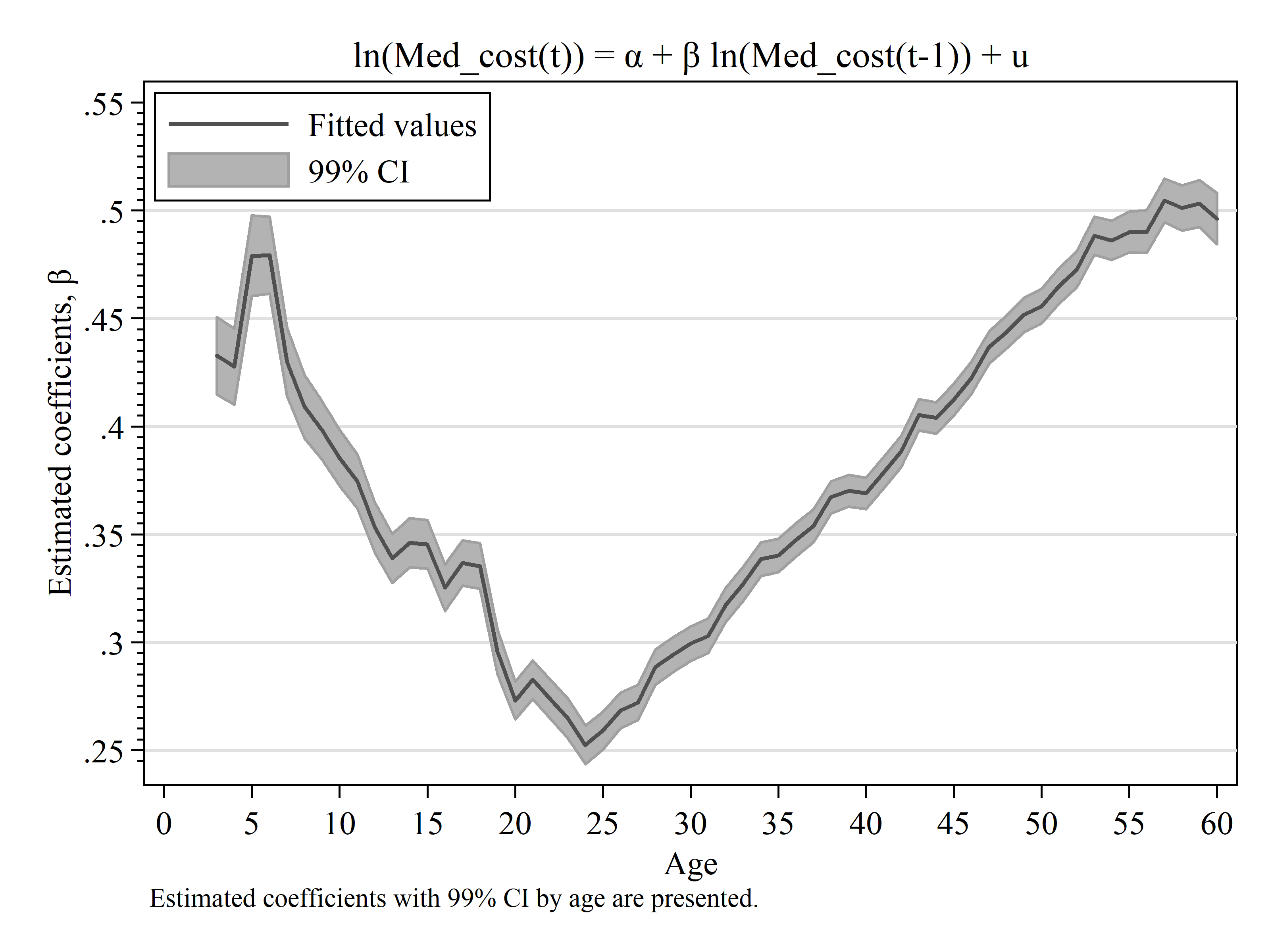}
	\caption{Medical Fee Persistency, AR(1)} 
	\source{The Japan Medical Data Center (JMDC) claim database}
	\label{k15}
\end{figure}

\begin{figure}[!ht]
	\centering
	\includegraphics[scale=0.11]{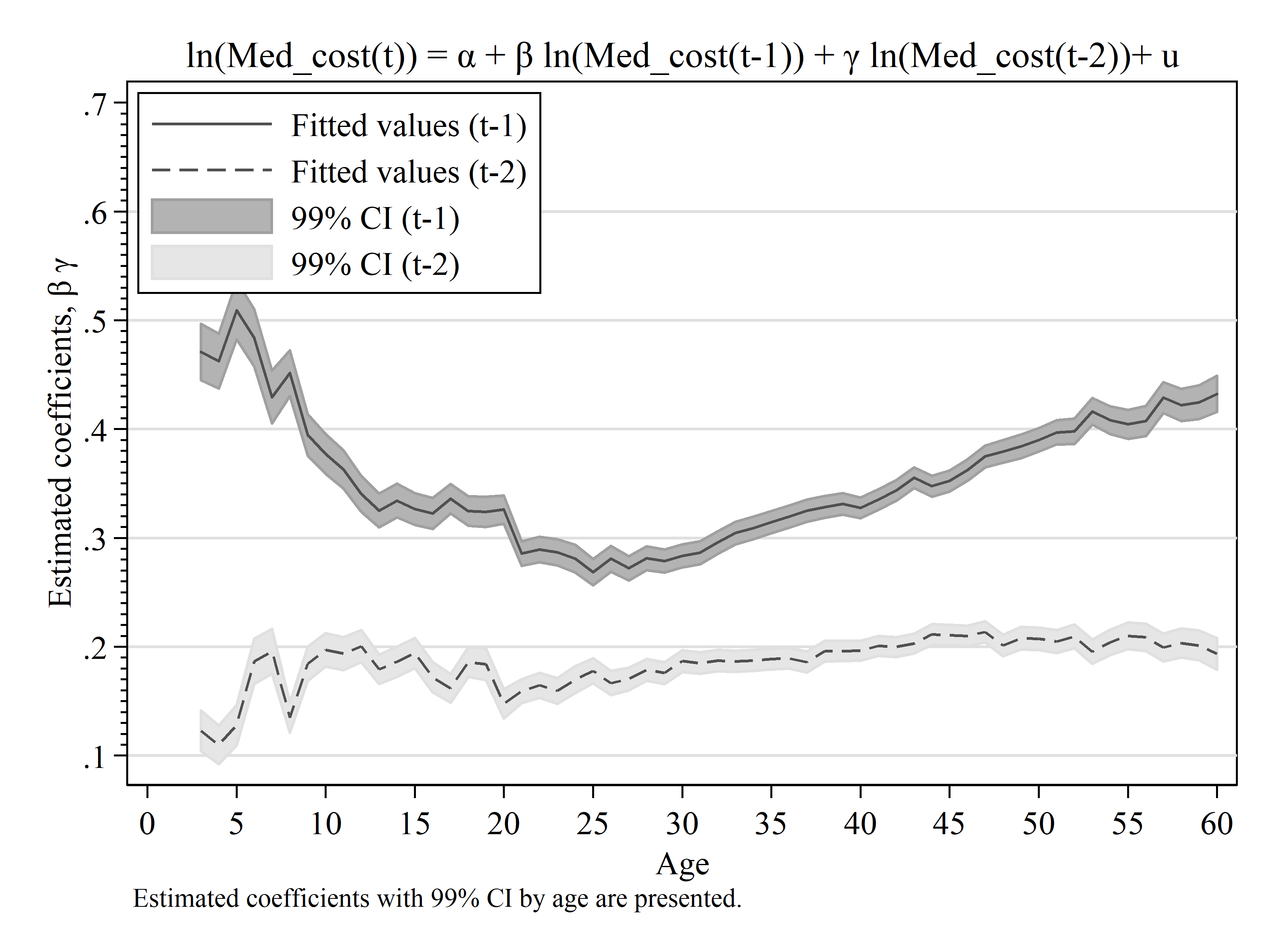}
	\caption{Medical Fee Persistency, AR(2)} 
	\source{The Japan Medical Data Center (JMDC) claim database}
	\label{k16}
\end{figure}

Figures \ref{k15} and \ref{k16} plot the simple auto regression results for health
expenditures for each age, including year dummies.  Similar to our
previous results, they show that the coefficients for the incurred medical fees
for the previous year, which corresponds to the persistency of a health
shock, as we have argued, vary across ages, recording the minimum
values when individuals are in their 20s, and increasing from there
onwards as age increases.  We obtain similar results even if we include
the medical cost for two years before (Figure \ref{k16}), or include more lag
variables. These additional results confirm that our main results were
not due to the definition of the thresholds for the states.

In conclusion, when people experience a health shock at some age, that
effect may persist in the following years.  The persistency rate,
however, differs across ages: it decreases during ages 0 to early 30s,
falling to the minimum when individuals are in their 20s and early 30s,
but increases starting from late 30s.  These results imply that
consumers' dynamic behaviors, such as saving or participating in
health insurance, should differ across ages not only because of their
remaining lifetime or family structure, but also because of the differences
in the probability of a health shock and in health shock persistency.

\subsection{Implications of the Results}
At the end, we conduct a brief simulation of the expected medical costs after health shocks by using the Markov chain of order two model.
In the previous section, we estimated the transition probability matrix for each age group.
By using these results, now, we can calculate the predicted medical costs by multiplying the sum of the incurred medical costs in each year with the transition probability of each state.
By repeating the same operation, we simulate the medical costs for the period of ten years after the initial health shock, i.e. from the best health status (state $Q1$) to the worst health state (state $Q5$), and then compare the predicted medical costs of those who experienced such health shocks and those who did not.\footnote{For more details, see the appendix.}
Note that we are looking at the ten-year total medical cost for which at least someone needs to pay.\footnote{Here we have focused not on individuals' health expenditure but the total medical cost mainly due to the following two reasons. First, we do not observe the actual expenditure when it comes to some of the individuals who have the poorest health status (state $Q5$). This is because some of them make use of the high-cost medical expense benefit system and some do not. In addition, some municipal governments offer free medical services to children through medical expenses subsidy programs aimed at providing support for child-rearing, so we are not able to precisely measure the actual medical expenditure for children either.}  
We choose the median of the threshold values of each state as the medical costs for that state, except for ($Q5$).
For state ($Q5$), which represents the poorest health, we change the value of the annual medical cost in order to investigate various possible cases.

\begin{figure}[!ht]
	\centering
	\includegraphics[scale=0.11]{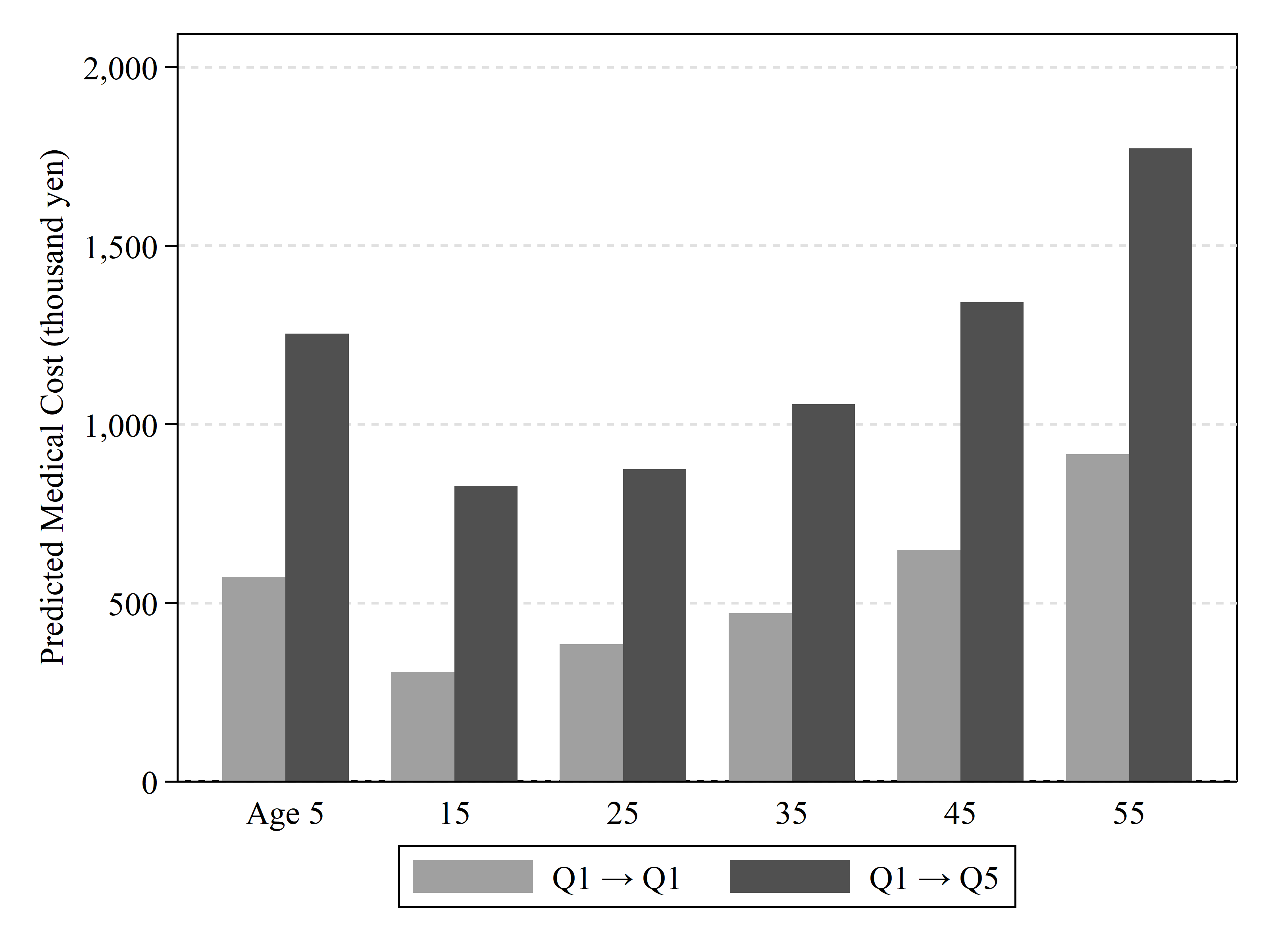}
	\caption{Predicted Ten-Year Medical Cost at Age $t$, $Q5$ = 267,000 yen} 
	\source{The Japan Medical Data Center (JMDC) claim database}
	\label{f02}
\end{figure}

We first examine the case in which individuals in the worst health state ($Q5$) incur the medical cost of 267,000 yen every year - the amount which comes from the threshold value of medical costs for those who are in the worst health state ($Q5$) (Figure \ref{f02}). 
Since we employed the lowest possible value for the worst health state ($Q5$), the predicted ten-year medical cost should represent the lower bound.
The simulation results show that, for all age groups, the predicted ten-year medical cost will almost double if individuals experience health shocks, i.e. the transition from the best health status ($Q1$) to the worst health status ($Q5$).
In addition, the differences in the predicted ten-year medical costs exhibit a U-shaped age profile, reflecting the fact that cost persistency differs according to age, as we have seen in section 6.3.
Also, even in this optimistic scenario, the ten-year total medical cost incurred by those who moved from the best health status (state $Q1$) to the worst health status (state $Q5$) at age 55 amounts to about 1.8 million yen.


Next, we consider two more cases: that individuals who are in poorest health (state $Q5$) incur the annual medical cost of (1) 0.5 million yen, and (2) one million yen.  
We first investigate how these two cases appear in our dataset, i.e., search for individuals who have moved from the best health status (state $Q1$) to the worst health status (state $Q5$), and check how many of them incurred the above-mentioned amounts of annual medical cost. 
The probabilities of encountering such major health shocks are presented in Table \ref{tab:table8}.
Among those who moved from the state of best health ($Q1$) to the state of worst health ($Q5$), about a half encountered a health shock that cost around 0.5 million.
The fraction of individuals whose annual medical costs amounted to one million yen was smaller than that of those who suffered a 0.5 million yen shock, about 20--30\%, but increased with age.

\begin{table}[htbp]
	\centering
	\caption{The Proportion of Subjects Encountering Medical Costs of above 500,000 and 1,000,000 Yen among Those Who Transitioned from Best Health (state $Q1$) to Poorest Health (state $Q5$) (\%)}
	\label{tab:table8}
	\begin{threeparttable}
		\begin{tabular}{cccc}
			\toprule
			\toprule
				  &		  & \multicolumn{2}{c}{Medical Costs above} \\
			   &       &  500,000 yen & 1,000,000 yen  \\
			   Age &    &  Proportion (\%) & Proportion (\%) \\
			\midrule
			0-4   &       & 34.2  & 8.2   \\
			5-9   &       & 41.9  & 12.5  \\
			10-14 &       & 47.6  & 15.4  \\
			15-19 &       & 55.7  & 25.0  \\
			20-24 &       & 53.0  & 21.8  \\
			25-29 &       & 50.2  & 21.2  \\
			30-34 &       & 50.0  & 20.2  \\
			35-39 &       & 50.3  & 22.0  \\
			40-44 &       & 53.4  & 24.9  \\
			45-49 &       & 55.1  & 28.5  \\
			50-54 &       & 59.8  & 32.5  \\
			55-59 &       & 61.9  & 36.2  \\
			\bottomrule
			\bottomrule
		\end{tabular}
	\begin{tablenotes}
		\small
		\item Data Sources: The Japan Medical Data Center (JMDC) claim database
		\item We defined the health transition states according to individual's overall medical cost for that year: 0--7,800 yen for $Q1$, 7,801--24,000 yen for $Q2$, 24,001--54,000 yen for $Q3$, 54,001--266,999 yen for $Q4$ and over 267,000 yen for $Q5$.
	\end{tablenotes}
	\end{threeparttable}
\end{table}

\begin{figure}[!ht]
	\centering
	\includegraphics[scale=0.11]{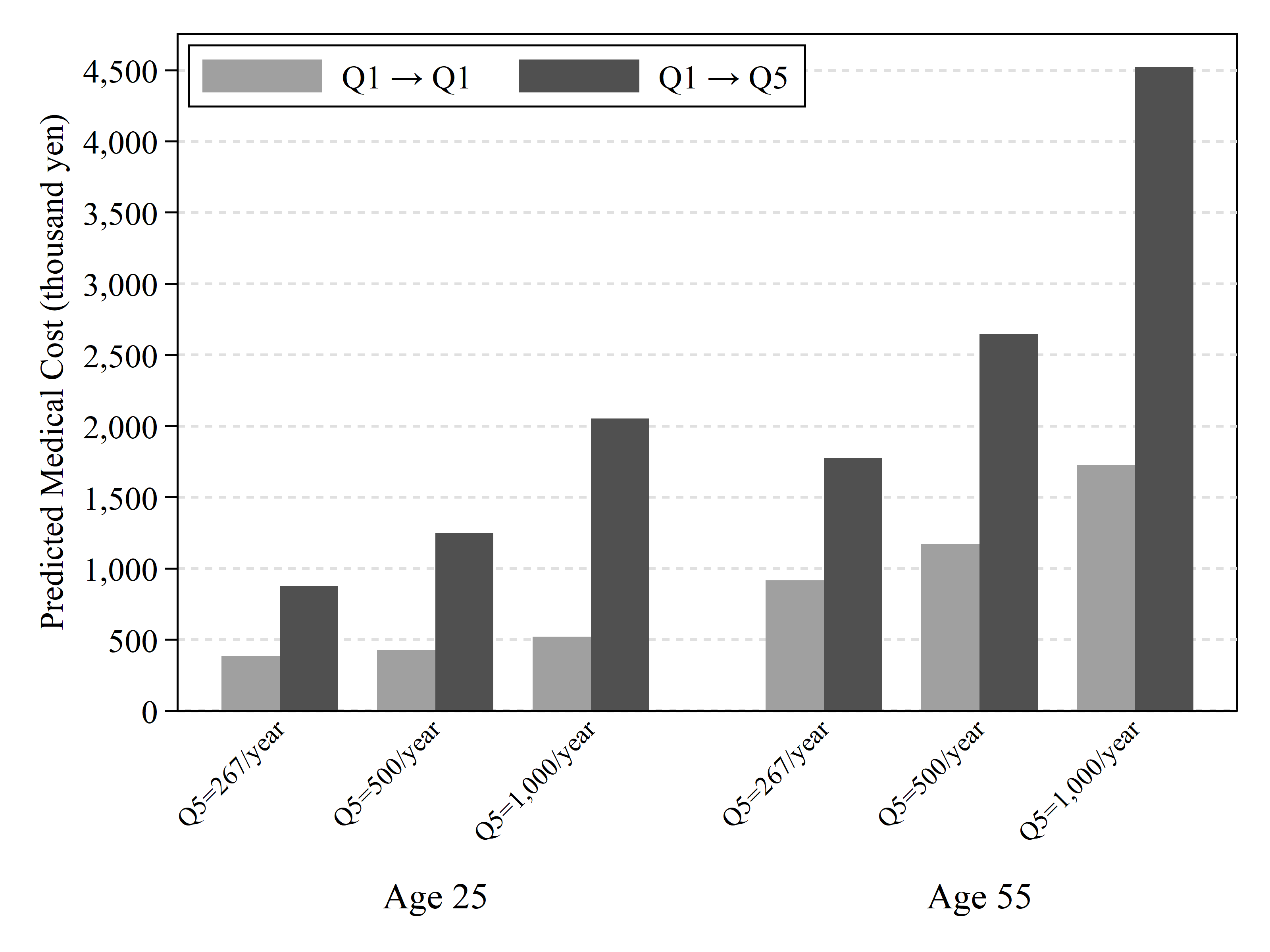}
	\caption{Predicted Ten-Year Medical Cost at Age $t$} 
	\source{The Japan Medical Data Center (JMDC) claim database}
	\label{f03}
\end{figure}

A new setting and the simulated medical costs of those who experience a health shock and those who do not is presented in Figure \ref{f03}. Here we focused only on the results for subjects age 25 and 55 for the sake of simplicity and comparison. 
As can be seen from the figure, the difference in medical expenditure over the period of ten years between those who suffered health shocks and those who did not increases.
While in the case of the lower threshold the difference was about 0.9 million yen for males aged 55, the difference grew to about 2.8 million yen in the case where the annual medical costs was one million yen.
In addition, the predicted ten-year medical cost of those who moved from the best health status (state $Q1$) to the worst health status (state $Q5$) increases significantly.
For example, the predicted ten-year medical cost for males aged 55 rises from 1.8 million yen to about 4.5 million yen, if we conduct simulation using the value of one million yen for the worst health status (state $Q5$).

To sum up, these simulation results indicate that once individuals transition from the best to the worst health status, their economic burden more than doubles over the following ten years.
In addition, taking into account that health insurance users in Japan are reimbursed a portion of the amount they paid through co-payment in case the medical cost that they had incurred was very high, these results also imply that the economic burden is non-negligible not only for the individuals, but also for the entire health insurance system, especially in the case of the cost generated by older men.


\section{Conclusion}
We have demonstrated the importance of using the Markov chain of order
two to capture health transitions appropriately.  By defining
health status in terms of absolute expenditure levels, rather than age-dependent
quantiles that are often used in the literature, we have traced the health transition
process over the life cycle for males.

In this paper, however, we have not utilized specific illnesses to classify the subjects. 
Conditioning based on such information, as well as the concrete treatment patients received would allow us to better
predict future outcomes.  Since patients possesses that type of information, it certainly affects their decisions, which is why it is important
to have more detailed physical information and conduct conditioning based on more precisely defined health transition states.

In this paper, following previous literature, we have taken health expenditure as the defining feature
of the health status. However, clearly, in reality health expenditure is an outcome determined jointly by underlying factors such as the 
health care system, the health insurance scheme and the economic conditions of households, 
as well as by health-related events in the life of an individual. 

For example, many municipalities in Japan offer free medical services to children
aged 15 or below through infants' and children's medical expenses subsidy programs.  
Thus, if health is defined in terms of medical costs or expenditure, individuals aged 16 or above may look healthier,
as they tend to abstain from using health care services more than children aged 15 or below, who have access to free medical services. 
Therefore, an important work that lies ahead of us is finding ways to overcome these difficulties.

\nocite{HSZ,SSK,DFJ,French,HJ,KS,Palumbo,NT,FJ,Hsu,PP1,PP2,JT,KK,HHL,Capatina,Yogo,AKV,Kotlikoff}

\singlespacing
\newpage
\bibliography{healthref}
\bibliographystyle{aer} 

\clearpage
\newpage
\doublespacing
\section*{Appendix}
\subsection*{Medical Fees Predicted by Using Markov Chain of Order Two}
We define the matrix of transition from $(i,j)$ into $(i^{\prime},j^{\prime})$,
where the first argument is the past state and the second argument
is the current state. In our case it is a $25\times25$ matrix. We order
the states so that initial states are in the order of 1, 2, 3, 4, 5 and
the ending states are in the same order, so that column $(i,j)$ is
ordered as $(1,1)$, $(1,2)$, $\ldots$, $(1,5)$, $(2,1)$, $(2,2)$,
$\ldots$, $(2,5)$, $\ldots$, $(5,1)$, $(5,2)$, $\ldots$, $(5,5)$
and the row is ordered as $(1,1)$, $(2,1)$, $\ldots$, $(5,1)$,
$(1,2)$, $(2,2)$, $\ldots$, $(5,2)$, $\ldots$, $(1,5)$, $(2,5)$,
$\ldots$, $(5,5)$.

We then compute the transition probabilities. The probability of moving
from $(i,j)$ to $(i^{\prime},j^{\prime})$ is zero if $j\neq i^{\prime}$,
and if $j=i^{\prime}$, then it equals
\[
\sum_{i=1}^{5}p(j^{\prime}|i,i^{\prime}).
\]
Let $\iota_{5}$ denote the vector of 5 ones and $e_{j}$ to denote
a vector with 1 as the $j$th element and 0 for the rest of the arguments.
Given this transition probability matrix $P$, the expectation in
the next period starting from state $j$ ($j=1,\ldots,25$) can be
computed as below, where $T$ denotes a transpose of a vector or a
matrix: let $M$ be a vector representing the health expenditure for
the 5 states $1,\ldots,5$,
\[
(M\otimes\iota_{5})^{T}Pe_{j}.
\]
The expectation in the second period is: 
\[
(M\otimes\iota_{5})^{T}P^{2}e_{j}.
\]
What we want to compute is the sum of these numbers for $e_{1}$ and
$e_{5}$, i.e. compare expectations for $(1,1)$ and $(1,5)$.

\end{document}